\documentclass[11pt, fullpage]{article}
\usepackage{fullpage}
\usepackage{setspace}
\usepackage{authblk}
\usepackage{cite}
\usepackage[colorlinks,citecolor=blue]{hyperref}
\usepackage[pdftex]{graphicx}
\usepackage{caption}
\usepackage{subfig}
\usepackage{amsmath}

\begin{document}
\title{ANN-assisted CoSaMP Algorithm for Linear Electromagnetic Imaging of Spatially Sparse Domains}
\author[1]{Ali I. Sandhu}
\author[1]{Salman A. Shaukat}
\author[2]{Abdulla Desmal}
\author[1]{Hakan Bagci}
\affil[1]{Division of Computer, Electrical, and Mathematical Science and Engineering (CEMSE) \newline King Abdullah University of Science and Technology (KAUST), Thuwal, Saudi Arabia, 23955-6900 \newline \{aliimran.sandhu, salman.shaukat, hakan.bagci\}@kaust.edu.sa}
\affil[2]{Department of Electrical Engineering, Higher Colleges of Technology, (HCT)\newline Ras Al-Khaimah, United Arab Emirates,  adesmal@hct.ac.ae}
\date{} 
\renewcommand\Affilfont{\itshape\small}

\maketitle
\onehalfspacing
\begin{abstract}
Greedy pursuit algorithms (GPAs) are widely used to reconstruct sparse signals. Even though many electromagnetic (EM) inverse scattering problems are solved on sparse investigation domains, GPAs have rarely been used for this purpose. This is because (i) they require a priori knowledge of the sparsity level in the investigation domain, which is often not available in EM imaging applications, and (ii) the EM scattering matrix does not satisfy the restricted isometric property. In this work, these challenges are respectively addressed by (i) using an artificial neural network (ANN) to estimate the sparsity level, and (ii) adding a Tikhonov regularization term to the diagonal elements of the scattering matrix. These enhancements permit the compressive sampling matching pursuit (CoSaMP) algorithm to be efficiently used to solve the two-dimensional EM inverse scattering problem, which is linearized using the Born approximation, on spatially sparse investigation domains. Numerical results, which demonstrate the efficiency and applicability of the proposed ANN-enhanced CoSaMP algorithm, are provided.  

\vspace{0.5cm}


\end{abstract}

\newpage
\doublespacing
\section{Introduction}\label{sec:introduction}
In the last two decades, greedy pursuit algorithms (GPAs) making use of compressed sensing (CS) have been successfully used  in reconstruction of sparse signals\cite{candes2008intro_compressive_sampling, signal_processing2011science, tropp2007signal, needell2009cosamp, needell2010signal}. In general, a CS-based scheme seek for the sparsest solution by constraining the optimization problem with the $\ell_{0}$-norm of the solution. It is well known that a direct solution of this optimization problem is NP-hard, however, GPAs, under certain conditions, provide a good approximation to its solution in an efficient way~\cite{tropp2007signal, needell2009cosamp, needell2010signal}. GPAs that are widely used in the image and signal processing communities rely on the orthogonal matching pursuit (OMP) \cite{tropp2007signal} and its different variants including the regularized OMP \cite{needell2010signal} and the compressive sampling matching pursuit (CoSaMP) \cite{needell2009cosamp}. 

Many electromagnetic (EM) imaging applications involve naturally sparse investigation domains. Here, sparsity means that the scatterer, which is defined by a contrast function (difference between the dielectric permittivity of the scatterer and the background medium), occupies only a small fraction of the investigation domain, i.e., contrast function is spatially sparse\cite{massa2015compressive}. Note that sparsity can also be defined with respect to a different basis expansion or in a transformed domain \cite{anselmi2015wavelet,li2013contrast, sandhu2016sparsity}. Having said that, application of GPAs to the solution of the EM inverse scattering problem has been rather limited~\cite{2008somp, 2012phaseless_imaging,2015ftb_omp}. This can be explained by two fundamental reasons: (i) For accurate and efficient reconstruction, GPAs require the exact number of non-zero elements (samples of the contrast function) in the investigation domain (denoted by $k$ in this work)  to be known a priori and this information is not available in many EM imaging applications. (ii) The GPAs require the measurement matrix to satisfy the restricted isometric property (RIP) \cite{tropp2007signal,needell2009cosamp,needell2010signal}. For EM imaging applications, the measurement matrix is the scattering matrix and is obtained by sampling the Green function of the background medium between the measurement/receiver locations and the investigation domain. This matrix does not satisfy the RIP~\cite{2008somp, 2012phaseless_imaging,2015ftb_omp}.

In~\cite{2008somp}, a simultaneous OMP algorithm (S-OMP) is developed to reconstruct sparse investigation domains, however, the algorithm is provided with the value of $k$, which would not be possible in many EM imaging applications. In~\cite{2012phaseless_imaging}, a phase-less reconstruction scheme is developed to reconstruct low-contrast point like scatterers. The optimization problem is linearized under the Born approximation~\cite{pastorino2010microwave} and made convex using $\ell_1$-norm regularization. In~\cite{2015ftb_omp}, a flexible tree search based OMP algorithm (FTB-OMP) is developed to reconstruct closely-spaced point-like scatterers and $k$ is predicted using the data misfit at each stage of the tree search. This scheme suffers from a tradeoff between the reconstruction accuracy and the computational cost, both of which increase with the search tree size. Another limitation of the FTB-OMP is that if the very first estimate of the solution component is incorrect, the algorithm would converge to a local minimum or an entirely incorrect solution.

It should be mentioned here that the EM inversion schemes making use of Bayesian compressive sensing (BCS) frameworks (some of which also operate under first-order linearization schemes such as Born and Rytov approximations) have been developed~\cite{oliveri2019, fouda2014, poli2014, BA2012microwave, oliveri2012}. Since these algorithms belong to the family of probabilistic inversion schemes, they are not discussed here. The readers are referred to~\cite{oliveri2019, fouda2014, poli2014, BA2012microwave, oliveri2012} and the references cited in these papers for details.

In this work, CoSaMP is used to solve the EM inverse scattering problem, which is linearized using the Born approximation, on two-dimensional (2D) spatially sparse domains. Unlike OMP, CoSaMP works by estimating multiple elements in the support set of the scatterer (i.e., indices of non-zero contrast samples) instead of one at a time and refines this support set iteratively. Consequently, it offers a faster rate of convergence and can reconstruct large connected scatterers, which is a limitation associated with the OMP-based algorithms. The two main challenges associated with the application of GPAs to EM imaging (as described above) are respectively addressed by (i) using an artificial neural network (ANN) to estimate $k$ and (ii) applying Tikhonov regularization to ensure that the RIP is satisfied. The ANN is trained beforehand using synthetic but noisy (scattered field) measurements on many investigation domains (with known $k$) which include various scatterers with different contrast levels and are discretized using different number of elements. This ANN is then used to predict $k$ given the fields scattered from an unknown domain. The Tikhonov ($\ell_2$-norm) regularization term is added to the diagonal elements of the scattering matrix to ensure that the RIP is satisfied. This enables the use of CoSaMP and reduces the sensitivity of the solution to the noise in the measurements.
 
The advantages of this Tikhonov and ANN-enhanced CoSaMP algorithm are threefold: (i) It is computationally efficient since the least squares problem, which is solved for unknown values of the contrast samples, is executed on a reduced set rather than the full investigation domain, (ii)  it does not require tuning of a thresholding parameter~\cite{desmal2014shrinkage}, and (iii) the reconstructed images are more accurate and sharper than those produced by smoothness promoting inverse algorithms.

It should be emphasized here that the ANN used in this work predicts only the number of non-zero elements ($k$) but neither their locations nor values. From this perspective, it can be considered as a ``qualitative’’ imaging approach and its structure is much simpler compared to the ANNs used for ``quantitative’’ imaging where the contrast is fully reconstructed~\cite{2018deepnis, sun2018efficient, bermani2000inverse, mccann2017convolutional, massa2018learning}. Having said that, it still comes with a computational overhead due to training (even though it is done only once) and one can extend other ``qualitative’’ methods \cite{litman1998reconstruction, dorn2006level, benedetti2010multiple, cakoni2006analysis, colton2003linear} to predict $k$ while avoiding this computational cost. 


It should also be noted here that a preliminary version of the CoSaMP framework proposed here has been described in a conference contribution~\cite{sandhu2017}. This inversion scheme work on current samples and is provided the exact value of $k$.

\section{Formulation}
\subsection{EM Inverse Scattering Problem} \label{sec:cs_formulation}
Let $S$ represent the support of a 2D inhomogeneous investigation domain that resides in an unbounded background medium. The permittivity and permeability in $S$ and in the background medium are $\{ \varepsilon ({\mathbf{r}}),{\mu _0}\} $ and $\{ {\varepsilon _0},{\mu _0}\} $, respectively. $S$ is individually illuminated by $N^{\mathrm{T}}$ line-source transmitters that generate transverse magnetic (to $z$) $\mathrm{TM}^z$ incident fields $E_i^{\mathrm{inc}}({\mathbf{r}})$, $i = 1, \ldots , N^{\mathrm{T}}$ at frequency $\omega = 2\pi f$. Upon excitation by $E_i^{{\mathrm{inc}}}({\mathbf{r}})$, an electric current density is induced on $S$ and this current density generates the scattered electric field $E_i^{{\mathrm{sca}}}({\mathbf{r}})$. Let $E_i^{{\mathrm{tot}}}({\mathbf{r}}) = E_i^{{\mathrm{inc}}}({\mathbf{r}}) + E_i^{{\mathrm{sca}}}({\mathbf{r}})$ represent the total electric field, then $E_i^{{\mathrm{sca}}}({\mathbf{r}})$ can be expressed as~\cite{pastorino2010microwave}:
\begin{equation} \label{eq1}
E_i^{{\mathrm{sca}}}({\mathbf{r}}) = {k_0^2}\int\nolimits_S {\tau ({\mathbf{r'}})E_i^{{\mathrm{tot}}}({\mathbf{r'}})G({\mathbf{r}}_m^{\mathrm{R}},{\mathbf{r'}})ds'}.
\end{equation} 
Here, ${k_0} = \omega \sqrt {{\varepsilon _0}{\mu _0}} $ is the wavenumber in the background medium, $\tau(\mathbf{r}) =\varepsilon(\mathbf{r})/\varepsilon_0-1$ is the dielectric contrast, and $G({\mathbf{r}},{\mathbf{r'}}) = H_0^2(k_0 \left| {\mathbf{r}} - {\mathbf{r'}}\right|)/(4j)$ is the 2D Green function.

\begin{figure}[b!]
\centerline{\includegraphics[width=0.4\columnwidth]{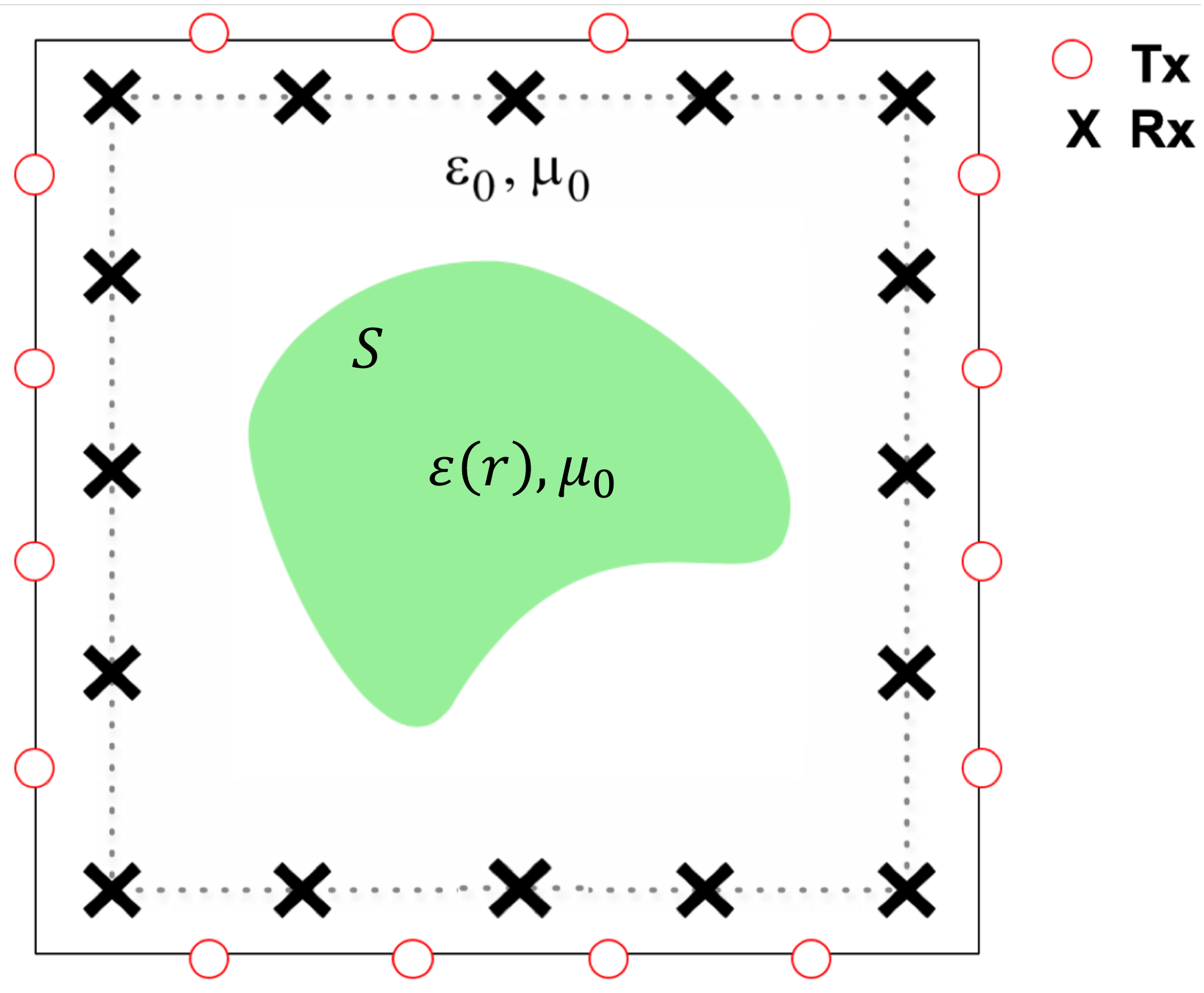}}
\caption{Description of the 2D EM inverse scattering problem.}
\label{fig1}
\end{figure}

The EM inverse scattering problem (Fig. \ref{fig1}) is defined as finding $\tau(\mathbf{r})$ given measured $E_i^{{\mathrm{sca}}}({\mathbf{r}_m^{\mathrm{R}}})$, where ${\mathbf{r}_m^{\mathrm{R}}}$, $m = 1, \ldots ,N^{\mathrm{R}}$ are receiver locations away from $S$. To solve this problem numerically, first, $S$ is discretized into $N$ number of square elements. Let $\mathbf{r}_n$, $n = 1, \ldots ,N$ represent the centers of these elements. Then, $E_i^{{\mathrm{tot}}}({\mathbf{r}})$ and $\tau ({\mathbf{r}})$ are approximated as 
\begin{align}\label{eq_exp}
E_i^{{\mathrm{tot}}}({\mathbf{r}}) &= \sum\nolimits_{n = 1}^N {{\{ \bar{E}_i^{\mathrm{tot}}\} _n}{p_n}({\mathbf{r}})}\\
\nonumber \tau ({\mathbf{r}}) & = \sum\nolimits_{n = 1}^N {{{\{ \bar \tau \} }_n}{p_n}({\mathbf{r}})} 	
\end{align}
where ${\{ \bar{E}_i^{\mathrm{tot}}\} _n} = E_i^{{\mathrm{tot}}}({{\mathbf{r}}_n})$, ${\{ \bar \tau \} _n} = \tau ({{\mathbf{r}}_n})$, and ${p_n}({\mathbf{r}})$ is the pulse basis function on element $n$ with support ${S_n}$. Note that ${p_n}({\mathbf{r}})$ is non-zero only for ${\mathbf{r}} \in {S_n}$ with unit amplitude. Substituting~\eqref{eq_exp}  into~\eqref{eq1} and evaluating the expression at ${\mathbf{r}_m^{\mathrm{R}}}$, $m = 1, \ldots ,N^{\mathrm{R}}$ yields a matrix equation
\begin{equation} \label{eq2}
\bar{E}_i^{\mathrm{sca}} = \bar{G}\bar{D}\{\bar{E}_i^{\mathrm{tot}}\}\bar{\tau}, \; i = 1, \ldots , N^{\mathrm{T}}
\end{equation} 
where ${\{ \bar{E}_i^{\mathrm{sca}}\}_m} = E_i^{{\mathrm{sca}}}({\mathbf{r}}_m^{\mathrm{R}})$, $\bar{\bar{D}}\{\bar{E}_i^{\mathrm{tot}}\}$ is a diagonal matrix with entries ${\{ \bar{E}_i^{\mathrm{tot}}\} _n}$, and the entries of the matrix ${\bar{G}}$ are ${\{ \bar G\}_{m,n}} = k_0^2\int_{{S_n}} {G({\mathbf{r}}_m^{\mathrm{R}},{\mathbf{r'}})ds'}$. 

Matrix equation \eqref{eq2} is nonlinear in $\bar \tau$ since $\bar{E}_i^{\mathrm{tot}}$ is a function of $\bar \tau$. Under the assumption of weak scattering, \eqref{eq2} can be linearized using the Born approximation $\bar{E}_i^{\mathrm{tot}} \approx \bar{E}_i^{\mathrm{inc}}$, where ${\{ \bar{E}_i^{\mathrm{inc}}\} _n} = E_i^{{\mathrm{inc}}}({{\mathbf{r}}_n})$~\cite{pastorino2010microwave}. Inserting this approximation into~\eqref{eq2} yields a linear matrix equation
\begin{equation} \label{eq_h}
\bar{E}_i^{\mathrm{sca}} = \bar {H_i} \bar{\tau}= \bar{G}\bar{D}\{\bar{E}_i^{\mathrm{inc}}\}\bar{\tau}, \; i = 1, \ldots , N^{\mathrm{T}}.
\end{equation} 
This equation is ``inverted'' for $\bar \tau$ using the CoSaMP algorithm~\cite{needell2009cosamp} as described in the next section. 


\subsection{CoSaMP for EM Inverse Scattering Problem} \label{sec:cosamp}
In this work, the CoSaMP algorithm~\cite{needell2009cosamp} is used to solve \eqref{eq_h} for sparse $\bar \tau$ as described next. First, \eqref{eq_h} is expressed as an optimization problem  
\begin{equation}\label{eq3} 
\bar{\tau} = \mathop {\min }\limits_{\bar{\tau}} {\left\| {\bar{\tau}} \right\|_0}  \qquad\mathrm{s.t.}   \left\| {\bar{E}^{\mathrm{meas}} - {\bar H}\bar{\tau}} \right\|_2^2 \leq \epsilon. 
\end{equation}
Here, $\bar{E}^{\mathrm{meas}}  \approx \bar{E}^{\mathrm{sca}} + \bar{\eta}$, where $\bar{\eta}$  represents the additive white Gaussian noise, and $\bar{E}^{\mathrm{sca}}$ and $\bar H$ are obtained by respectively cascading $\bar{E}_i^{\mathrm{sca}}$ and $\bar H_i$ for all $i = 1, \ldots , N^{\mathrm{T}}$. 

For the CoSaMP algorithm to converge, the scattering matrix ${\bar H}$ has to satisfy the RIP\cite{needell2009cosamp}. This condition states that for any vector $\bar{y}$  there should be $\delta\in(0,1)$, such that $(1-\delta)\|\bar{y}\|^2_2 \leq \|{\bar H}\bar{y}\|^2_2 \leq (1+\delta)\|\bar{y}\|^2_2$  holds. The largest value of $\delta$ that satisfies the RIP is known as the restricted isometric constant (RIC). For any system that satisfies the RIP with RIC value $\delta_r$, the following condition holds \cite{needell2009cosamp}:
\begin{equation}\label{eq4} 
\sqrt{1-\delta_r} \leq \sigma({\bar H})_{\mathrm{min}} \leq \sigma({\bar H})_{\mathrm{max}} \leq \sqrt{1+\delta_r}.  
\end{equation}
Here, $\sigma({\bar H})_{\mathrm{min}}$ and $\sigma({\bar H})_{\mathrm{min}}$ are the largest and smallest singular values of $\bar{H}$. For the EM inverse scattering problem, the lower bound in~\eqref{eq4} is not satisfied since $\bar H$ is ill-conditioned with $\sigma({\bar H})_{\mathrm{min}} \to 0$, which enforces $\delta_r \to 1$, and consequently results in violation of the RIP. To alleviate this problem and enable the use of CoSaMP in solving the optimization problem~\eqref{eq3}, the data misfit $ || {\bar{E}^{\mathrm{meas}} - {\bar H}\bar{\tau}} ||$ is replaced with its Tikhonov-regularized version as
\begin{equation}\label{eq5} 
\bar{\tau} = \mathop {\min }\limits_{\bar{\tau}} {\left\| {\bar{\tau}} \right\|_0}  \quad\mathrm{s.t.}   \left\| {\tilde{\bar{E}}^{\mathrm{meas}} - {{\bar H}}^{\lambda}\bar{\tau}} \right\|_2^2   \leq \epsilon. 
\end{equation}
Here, $\tilde{\bar E}^{\mathrm{meas}} = {\bar H}^{\dagger}\bar E^{\mathrm{meas}}$, ${\bar H}^{\lambda} = {\bar H}^{\dagger}{\bar H} + \lambda \bar{\bar I}$, ${\bar H}^{\dagger}$ is the complex conjugate of ${\bar H}$, and $\lambda$ is a positive real constant. Note that \eqref{eq5} can also be written as 
\begin{equation}\label{eq6} 
\bar{\tau} = \mathop {\min }\limits_{\bar{\tau}} {\left\| {\bar{\tau}} \right\|_0}  \quad\mathrm{s.t.}   \left\| {\tilde{\bar{E}}^{\mathrm{meas}} - {{\bar H}}^{\dagger}{\bar H}\bar{\tau}} \right\|_2^2  + \lambda^2 {\left\| {\bar{\tau}} \right\|_2^2} \leq \epsilon.  
\end{equation}
The form of the optimization problem provided in \eqref{eq6} clearly shows that it is not only constrained by the sparsity requirement but also regularized using the $\ell_2$-norm of the solution weighted with $\lambda^2$. Note that $ \sigma({\bar H}^{\lambda})=\sigma({{\bar H}}^{\dagger}{\bar H}) + \lambda$, and since ${{\bar H}}^{\dagger}{\bar H}$ is ill-conditioned with $\sigma({{\bar H}}^{\dagger}{\bar H})_{\mathrm{\mathrm{min}}} \to 0$, $\sigma({{{\bar H}}^\lambda})_{\mathrm{\mathrm{min}}} \to \lambda$. Consequently, since $\lambda$ is a positive real constant, the lower bound of the RIP can now be satisfied with $\delta_r \in (0,1)$. 

The CoSaMP algorithm is applied to the optimization problem~\eqref{eq6} to yield a sparse solution. The steps of this algorithm read:
\begin{align*}
&{\rm Step}\;1:\; {\rm initialize}\; \bar{r}^{(0)} = {\bar E}^{\mathrm{meas}}, n = 0, k, \lambda \\ 
&{\rm Step}\;2:\; \rm repeat \\
&{\rm Step}\;2.1: \; n = n+1\\
&{\rm Step}\;2.2: \; \bar{y}^{(n)} =  {\bar{H}}^{\dagger}\bar{r}^{(n-1)} \\
&{\rm Step}\;2.3: \; \Omega^{(n)} = \mathrm{supp}_{2k}(|\bar{y}^{(n)}|)\\
&{\rm Step}\;2.4: \; F^{(n)} = \Omega^{(n)} \cup \Gamma^{(n)}\\
&{\rm Step}\;2.5: \; {\rm solve}\; ({\bar H}_{:,F^{(n)}}^{\dagger} {\bar H}_{:,F^{(n)}} + \lambda\bar{I} )\bar {s}^{(n)} = \tilde{\bar E}^{\mathrm{meas}}\\
&{\rm Step}\;2.6: \Gamma^{(n)} = \mathrm{supp}_{k}(|\bar s^{(n)} |), \; \bar{\tau}^{(n)}_{\Gamma^{(n)}}=\bar{s}^{(n)}, \; \bar{\tau}^{(n)}_{\Gamma_c^{(n)}}=\bar{0}\\
&{\rm Step}\;2.7: \; \bar{r}^{(n)} = {\bar E}^{\mathrm{meas}} - {\bar H}\bar{\tau}^{(n)}
\end{align*}
At Step $1$, several parameters are initialized. The number of non-zero entries in $\bar \tau$ is denoted as $k$ and predicted by providing ${\bar E}^{\mathrm{meas}}$ to an ANN trained as described in the next section. As explained above, the parameter $\lambda$ is primarily used to ensure that the RIP is satisfied. But it also smoothens the high frequency components since it acts as a Tikhonov-type regularizer. Therefore, it is selected heuristically to increase the robustness of the solution to noise while maintaining its sharpness within the support set of the scatterer (see below).  At Step $2.2$ the residual $\bar{r}^{(n-1)}$ from the last iteration is projected onto the model subspace to determine which components of the unknown model are yet to be determined. At Step $2.3$, function $\mathrm{supp}_L(.)$ function choses indices of the largest $L$ entries of its input vector. As a result, index/support set $\Omega^{(n)}$ stores $2k$ column-indices from ${\bar H}$ which contribute maximally towards the projection at Step $2.2$. At Step $2.4$, the newly identified support set $\Omega^{(n)}$ is unified with the final support set $\Gamma^{(n)}$ from Step $2.6$ (of the previous iteration) to eliminate any repetitions in the support elements. At Step $2.5$, $\bar s^{(n)}$ is computed by solving a least squares problem over the merged support set $F^{(n)}$. Note that ${\bar H}_{:,F^{(n)}}^{\lambda} = {\bar H}_{:,F^{(n)}}^{\dagger} {\bar H}_{:,F^{(n)}} + \lambda\bar{I}$ contain only the columns of ${\bar H}$ whose indices are in the merged support set $F^{(n)}$. This significantly reduces the computational cost in contrast to solving the least squares problem involving the ``whole'' matrix ${\bar H}$. At Step $2.6$, support set $\Gamma^{(n)}$ stores the indices of the largest $k$ entries of $|\bar s^{(n)}|$. These are stored in the correct entries of $\bar \tau^{(n)}$ while its remaining entries are set to zero. Finally, at Step $2.7$, $\bar{r}^{(n)}$ is updated. The algorithm is set to terminate if the residual between successive iterations does not change significantly, i.e. $ || \bar{r}^{(n)} - \bar{r}^{(n-1)} ||/|| \bar{r}^{(n)} || \leq 10^{-6}$.

\subsection{Sparsity Estimation} \label{sec:sparsity_estimation}
Greedy algorithms (including CoSaMP) are very efficient in reconstructing sparse signals however their reconstruction accuracy degrades if the number of non-zero elements (denoted as $k$ in the previous section) is not known a priori. Unfortunately, this is often the case for a typical EM inverse scattering problem.  In this work, an ANN, which is constructed, trained, and tested as described below, is used to predict $k$. The value of this prediction is denoted by $\hat{k}$. 

\begin{figure}[t!]
\centering
\subfloat[]{\includegraphics[width=0.40\columnwidth]{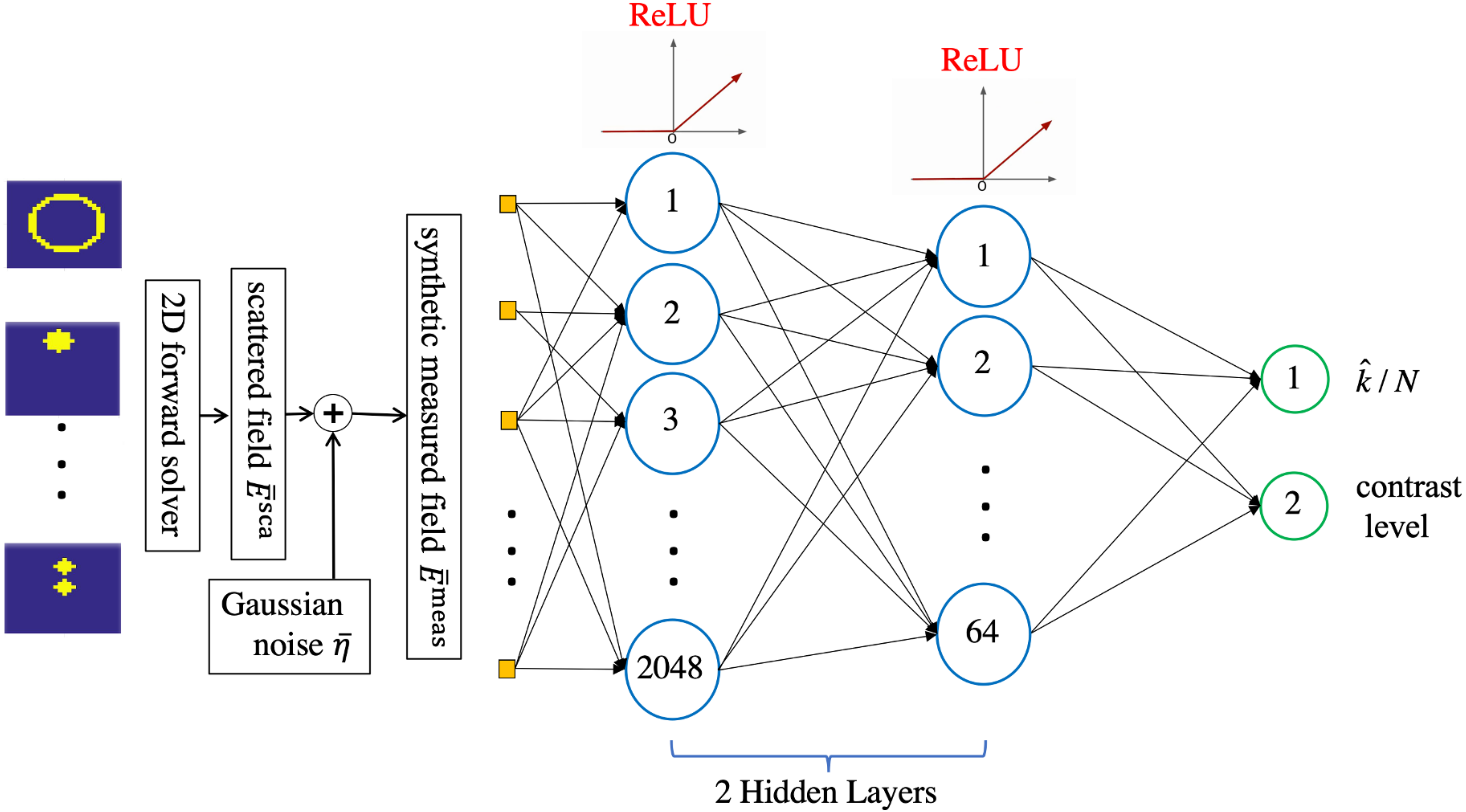}}
\subfloat[]{\includegraphics[width=0.40\columnwidth]{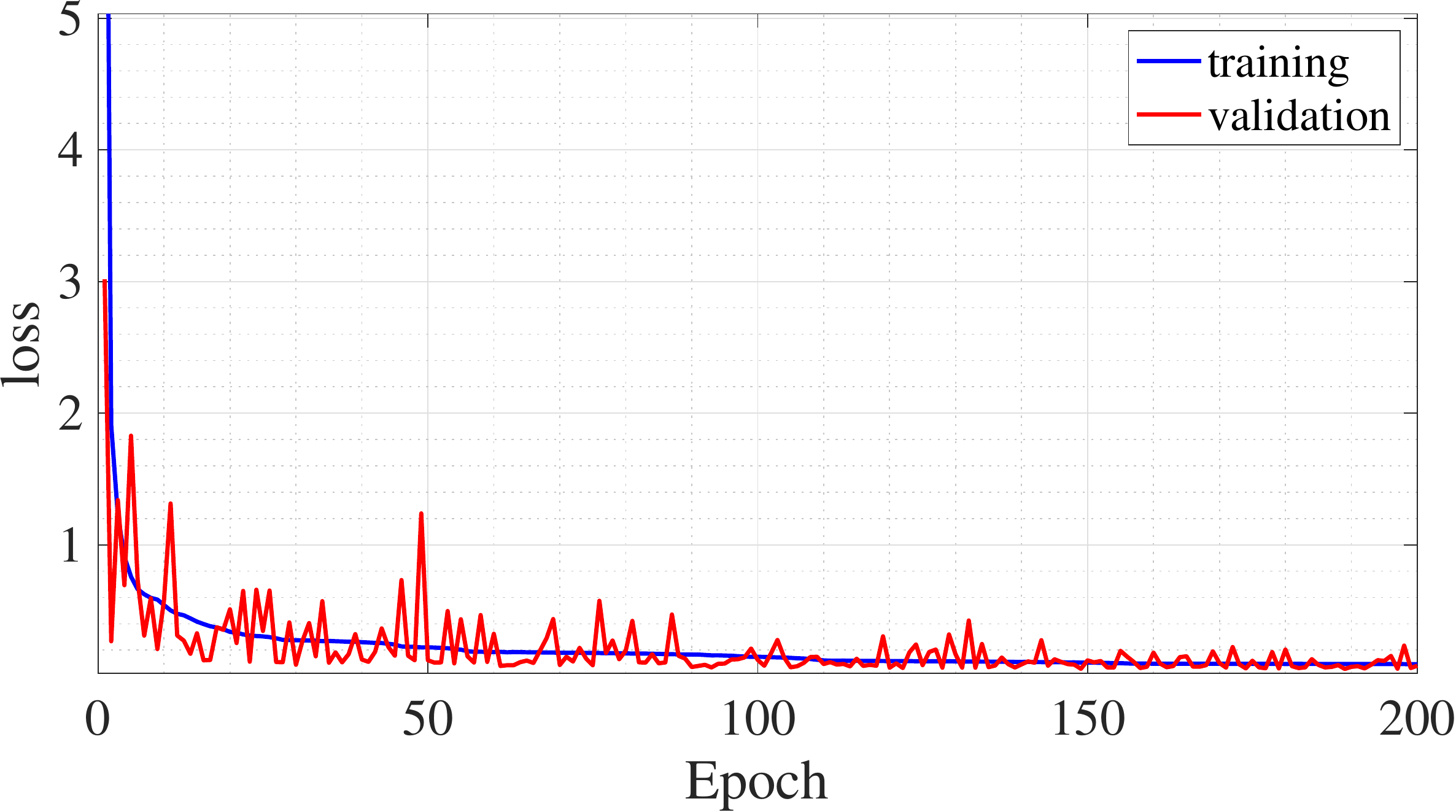}}\\
\vspace{-0.3cm}
\subfloat[]{\includegraphics[width=0.38\columnwidth]{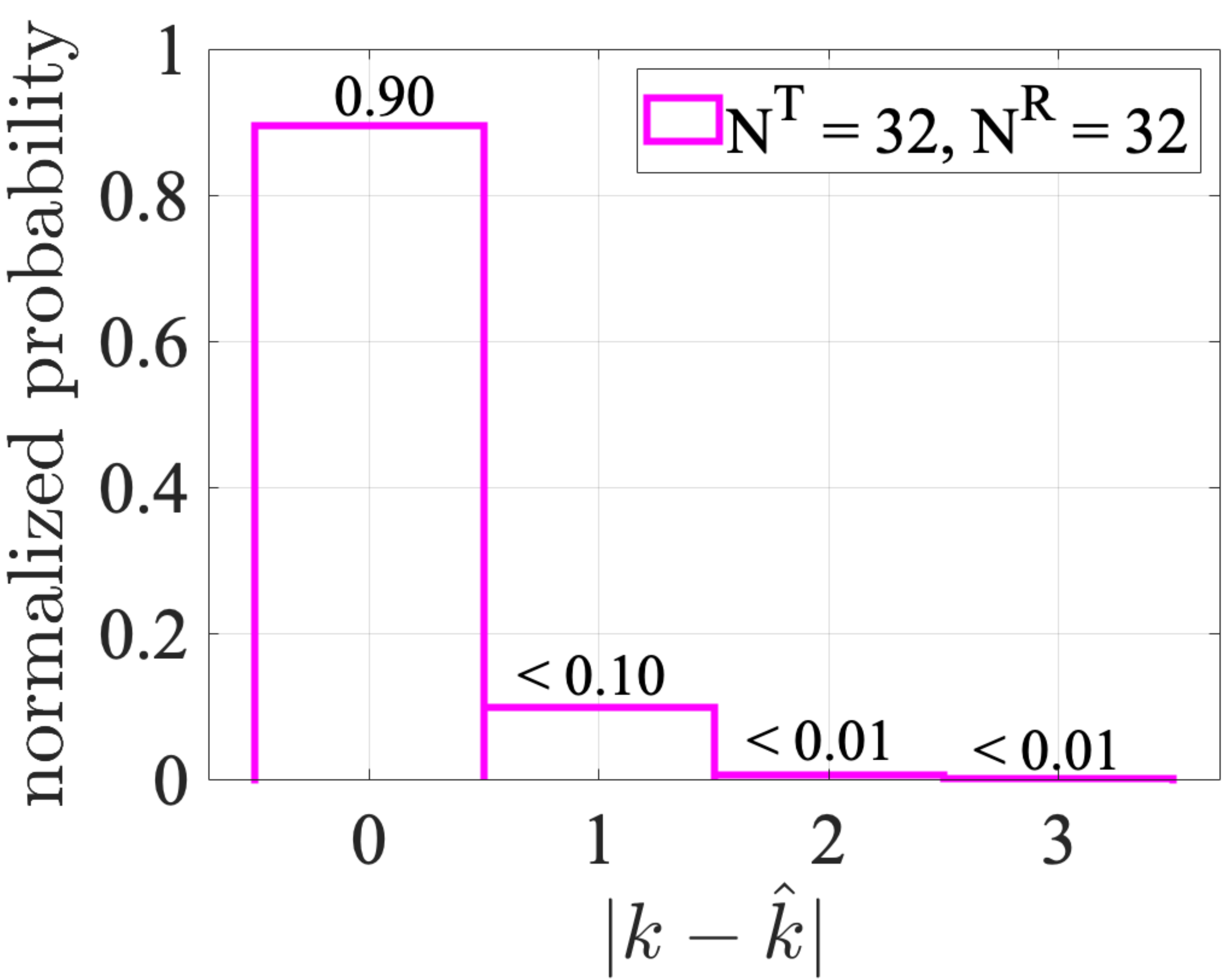}}
\subfloat[]{\includegraphics[width=0.38\columnwidth]{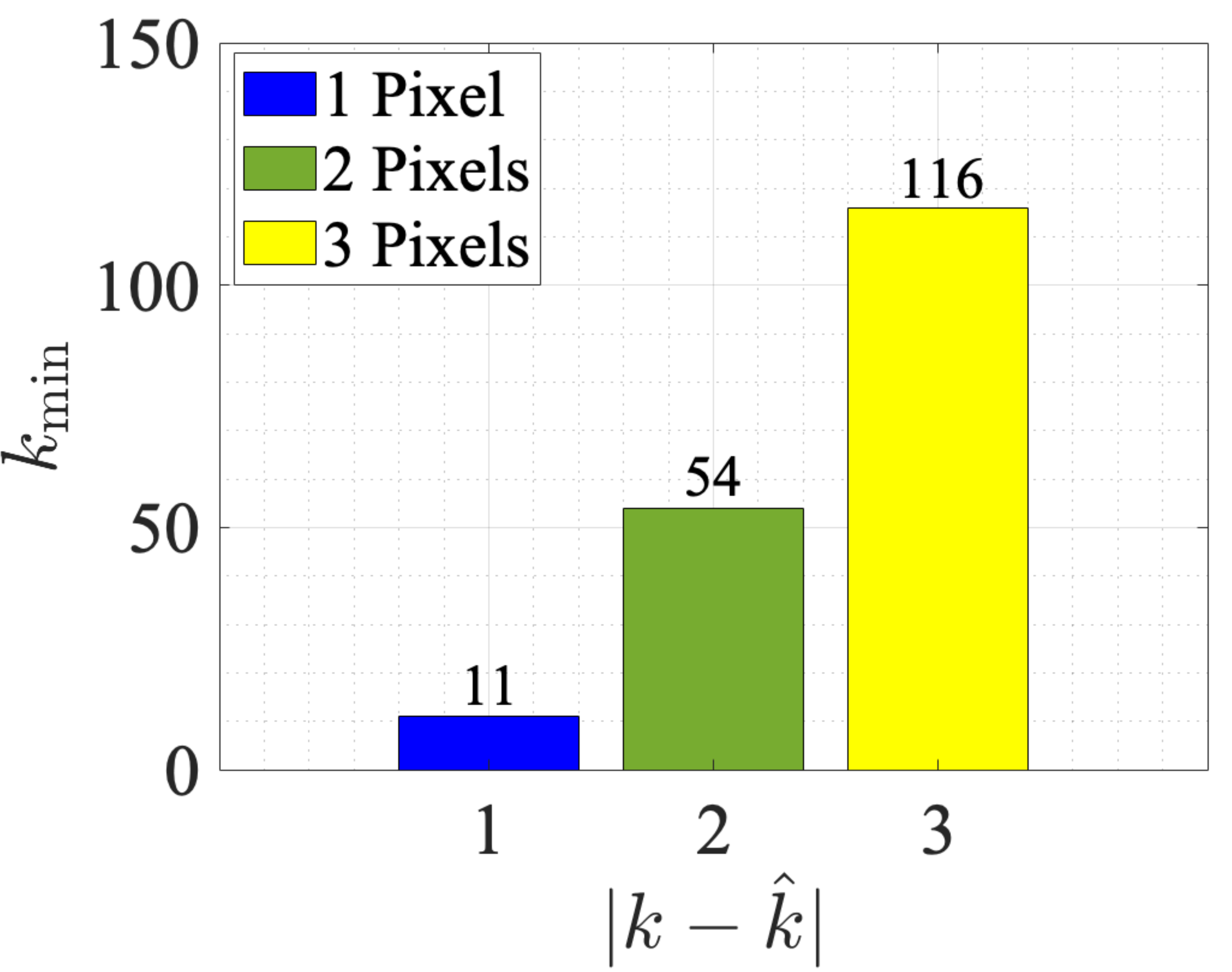}}\\
\vspace{-0.3cm}
\subfloat[]{\includegraphics[width=0.38\columnwidth]{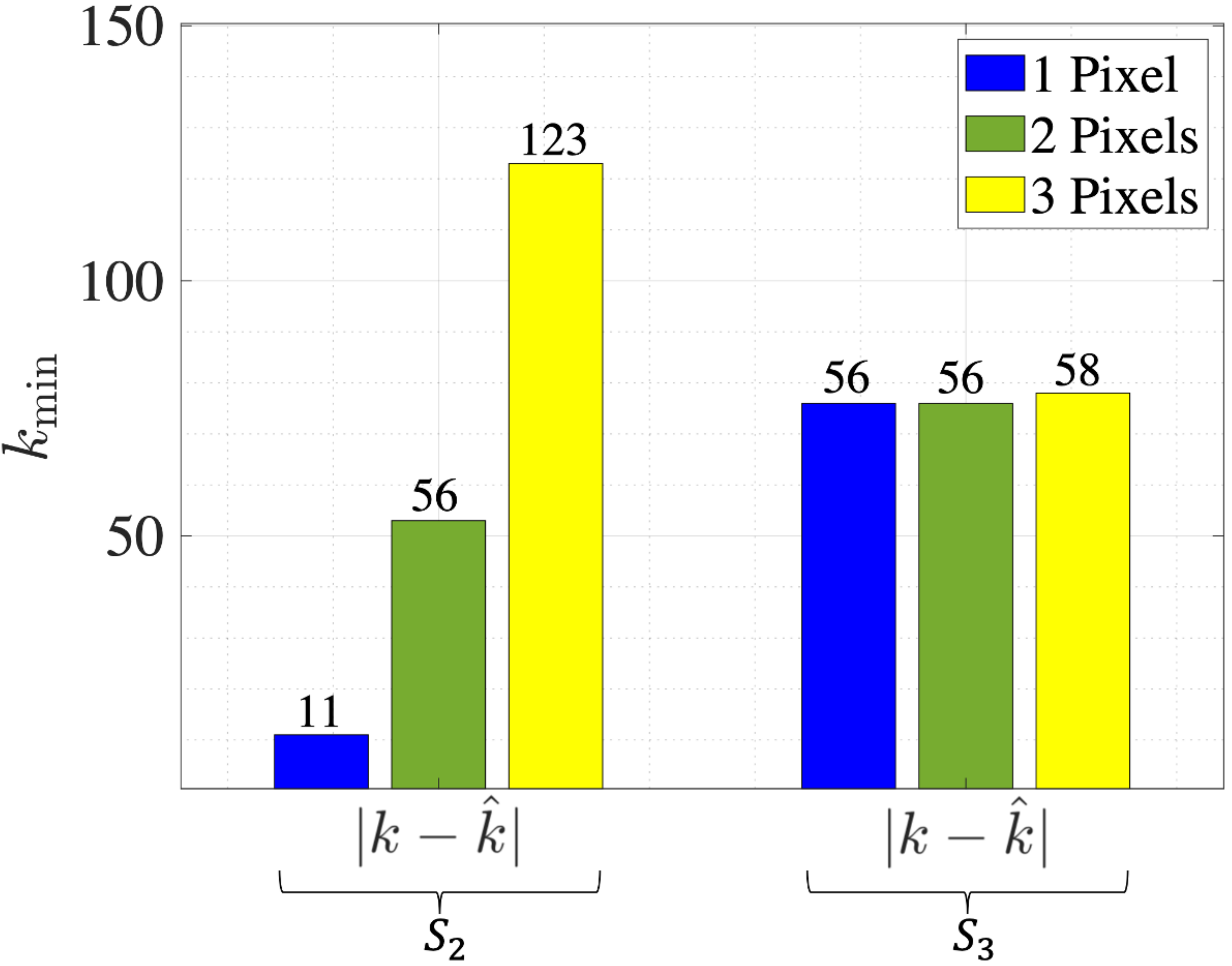}}
\subfloat[]{\includegraphics[width=0.42\columnwidth]{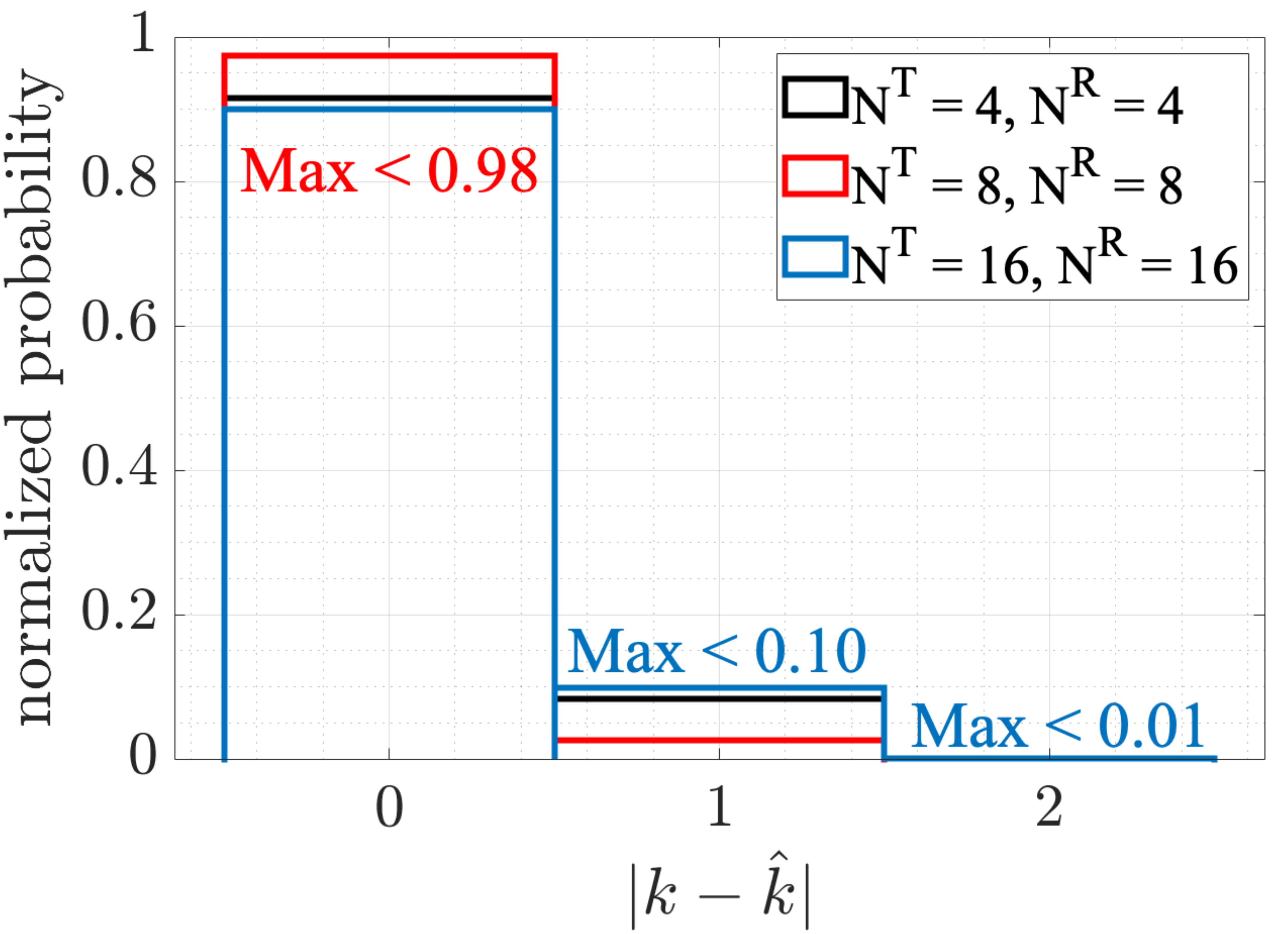}}\\
\vspace{-0.3cm}
\subfloat[]{\includegraphics[width=0.40\columnwidth]{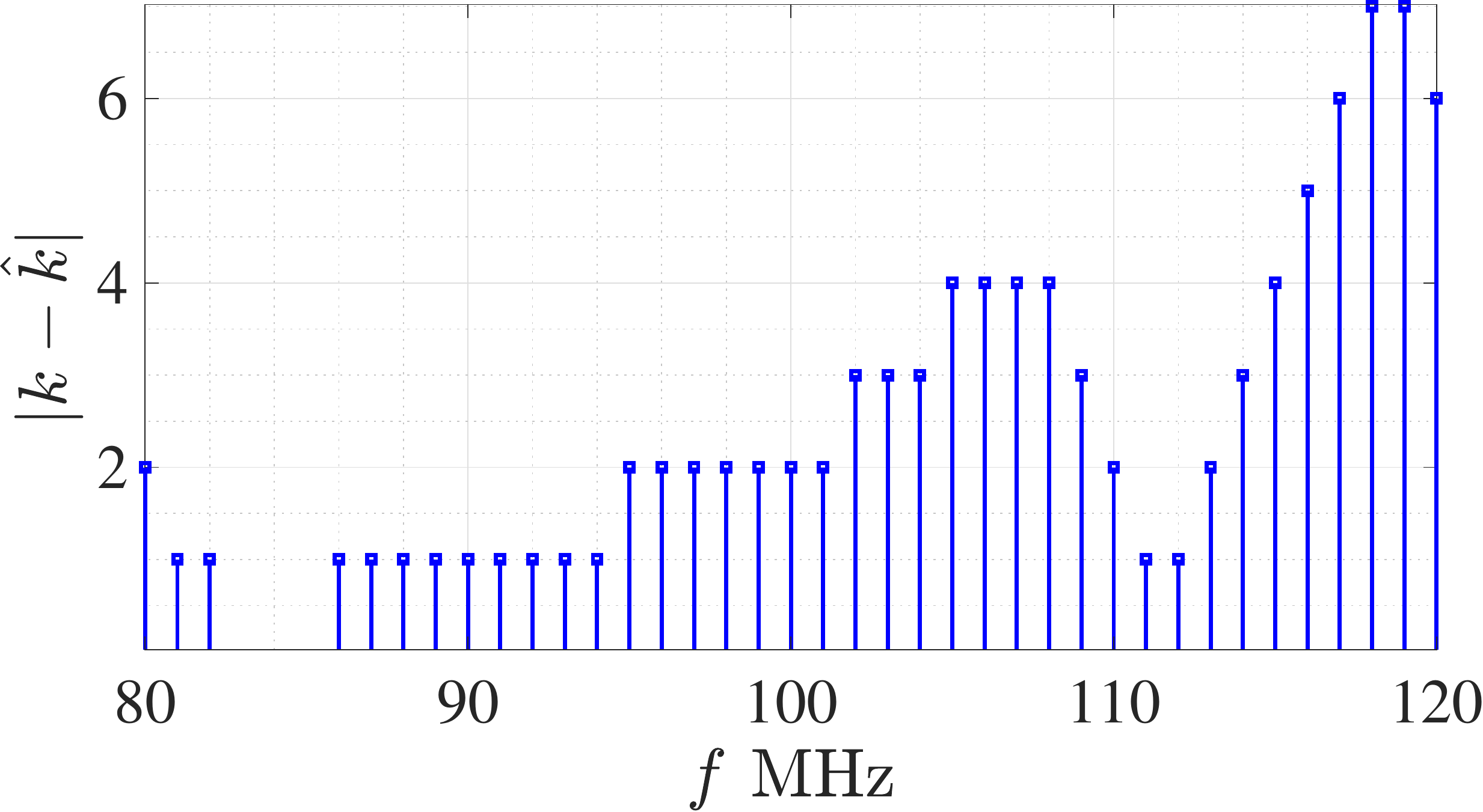}}
\vspace{-0.1cm}
\caption{ (a) ANN used to estimate the sparsity level $k$. (b) Convergence of the loss function for ANN-1 on $T_1$. (c) Normalized probability histogram of $|k-\hat k|$ obtained by ANN-1 over $S_1$. (d) $k_{\rm min}$ versus $|k-\hat k|$ over $S_1$ for ANN-1. (e)  $k_{\rm min}$ versus $|k-\hat k|$ over $S_2$ and $S_3$ for ANN-1. (f) Normalized probability histogram of $|k-\hat k|$ obtained by ANN-2, ANN-3 and ANN-4 (trained on $T_2$) over $S_4$. (f) $|k-\hat k|$ versus the frequency of a test scenario for ANN-1.}
\label{fig2}
\end{figure}

The ANN used here is a simple two-layer feedforward perceptron network~\cite{purkait2019keras}. As shown in Fig. \ref{fig2}(a), the two hidden layers consist of 2048 and 64 neurons. ``ReLU’’ is the activation function applied to the outputs of each layer~\cite{purkait2019keras,kochenderfer2019algorithms}. The training step minimizes a loss function that is defined as the mean squared error (MSE) between (i) $\hat{k}$ and $k$ (both of which are normalized with $N$) and (ii) the true and estimated contrast level. Only homogenous scatterers are used to train the ANN, i.e., contrast profile have the same value for all elements that represent the scatterer. This value is referred to as ``contrast value'' in the rest of the text. Note that the contrast level is not the reconstructed image. It is estimated at the ANN output and is used only to train the ANN better, which helps in producing a more accurate $\hat{k}$. This information is not used by the CoSaMP algorithm to reconstruct the contrast profile. The optimization algorithm is ``RMSprop’’ with a learning rate of 0.001 and default momentum of 0.0~\cite{kochenderfer2019algorithms}. 

This ANN directly accepts the scattered electric field measured/sampled at the receiver locations (together with added noise) at its input layer. In all training scenarios, the frequency of the transmitters $f=100$ MHz and A 2D volume integral equation solver (for $\mathrm{TM}^z$ fields)~\cite{pastorino2010microwave}, which uses the discretization scheme described by~\eqref{eq_exp} in the previous section, is used to compute the scattered field samples at the receiver locations. Then, $25\mathrm{dB}$ of Gaussian noise is added to these samples to generate a set of measurements. Four different ANNs are generated. For the first ANN, ANN-1, a total of $14450$ scattering scenarios including uniformly distributed scatterers (i) of different shapes, in particular, circular rings with random radii, single and double cylinders with random radii, and with varied separation distance, (ii) with contrast level selected from set $\{0.2, 0.4, 0.6, 0.8\}$, and (iii) with varying total number of discretization elements $N\in\{784,3136,12544\}$, are created. A randomly selected 70\% of this set is used to train the ANN, and the remaining 30\% is used to test it. These training and test sets are denoted by $T_1$ and $S_1$, respectively. For ANN-1, $N^{\rm T}=32$ and $N^{\rm R}=32$.

Fig.~\ref{fig2}(b) plots the convergence of the loss function for ANN-1 on $T_1$. Even though the contrast level is not used in the reconstruction by the CoSaMP algorithm, for the sake of completeness, the mean and variance of the error in the contrast level estimated by ANN-1 in $S_1$, are provided here: $0.088$ and $0.007$. Fig.~\ref{fig2}(c) shows the normalized probability histogram of the difference $|k-\hat{k}|$ obtained over $S_1$. It is observed that for $90\%$ of the test cases $|k-\hat{k}| = 0$ and the maximum value of $|k-\hat{k}|$ is $3$, which occurs for less than $1\%$ of the test cases in $S_1$. The histogram in Fig.~\ref{fig2}(c) presents only $|k-\hat{k}|$ without providing any information about the actual value of $k$. This figure is complemented by Fig.~\ref{fig2}(d)-(e) that plots $k_{\rm min}$ vs $|k-\hat{k}|$. Here, $k_{\rm min}$ is the smallest number of non-zero elements in the investigation domains of the test cases that produce the given value of $|k-\hat{k}|$. For example, in Fig. \ref{fig2}(d), for $|k-\hat{k}|=1$,  $k_{\rm min}=11$, which means that the largest relative error is $|k-\hat{k}|/ k_{\rm min} = 9\%$, and for $|k-\hat{k}|=3$, $k_{\rm min}=116$ and the largest relative error is $2.5\%$.

Two more test sets are generated to demonstrate the accuracy of ANN-1. Test set $S_2$ is created by selecting the scattering scenarios with $N = 784$ from $S_1$ and changing their contrast level in $\{0.3,0.7\}$. Note that these contrast levels are not used in $T_1$ which is used to train ANN-1. Test set $S_3$ is created using Austria-like profiles~\cite{van2001contrast} (also see Section III-C), which do not exist in $T_1$. $S_3$ includes a total of $400$ scattering scenarios where the scatterers include (i) a circular ring with constant radius and the two outer cylinders rotated around for $360^{0}$ in steps of $1^0$ (resulting in $360$ examples), and (ii) circular rings with varied radii and outer cylinders rotated in steps of $90^{0}$ (resulting in $40$ examples). The contrast level of all scatterers is set to $0.2$ and $N \in \{3136,12544\}$. The maximum value of $|k-\hat{k}|$ is 3 over all test examples in $S_2$ and $S_3$. Fig. \ref{fig2}(d) plots $k_{\rm min}$ versus $|k-\hat{k}|$ for both sets demonstrating the performance of ANN-1 for predicting the sparsity level of Austria-like profiles and when the test scatterers have contrast levels different from those of the training scatterers.

For ANN-2, ANN-3, and ANN-4, $\{N^{\rm T}, N^{\rm R} \} = $ $\{16,16\}$, $\{8,8\}$, and $ \{4, 4\}$, respectively. All three ANNs are trained on set $T_2$ which is created by selecting the scattering scenarios with contrast level $0.4$ and $N = 3136$ from $T_1$, and they are tested on set $S_4$ which is created from $S_1$ in the same way. Fig. \ref{fig2}(e) shows the three normalized probability histograms of the difference $|k-\hat{k}|$ obtained over $S_4$ using ANN-2, ANN-3, and ANN-4. The figure shows that the ANN is accurate in predicting $\hat{k}$ even when $N^{\rm T}$ and $N^{\rm R}$ are reduced to 4.

One last test is carried out to demonstrate the dependence of the prediction accuracy on the frequency. A test case with $N=12544$, $k=288$, and contrast level of $0.4$ is selected from $S_1$. ANN-1 (which is trained at $100$ MHz) is used to compute $\hat{k}$ while the frequency of the test scattering scenario is changed from $80$ MHz to $120$ MHz. Fig.~\ref{fig2}(f) plots $|k-\hat{k}|$ versus the frequency. The figure shows that the prediction accuracy deteriorates at higher frequencies but ANNs can still be used to provide $k$ for the CoSAMP algorithm at the frequencies that are closer to the frequency at which they are trained at [see Fig.~\ref{fig4}(d) and~\ref{fig5}(d)].

\section {Numerical results}  \label{sec:numerical_results}
This section demonstrates the accuracy and efficiency of the proposed scheme via numerical experiments. In all experiments, ${{\tau }^{\mathrm{ref}}}(\mathbf{r})$ and ${{\{{{\bar{\tau }}^{\mathrm{ref}}}\}}_{n}}={{\tau }^{\mathrm{ref}}}({{\mathbf{r}}_{n}})$, $n=1,\ldots ,N$ are the actual contrast and its samples, respectively. A 2D volume integral equation solver (for $\mathrm{TM}^z$ fields)~\cite{pastorino2010microwave}, which the discretization described in Section II-A is used to compute $\bar{E}^{\mathrm{sca}}$ from ${{\bar{\tau }}^{\mathrm{ref}}}$, then $\bar{E}^{\mathrm{meas}}$ is generated by adding Gaussian noise to $\bar{E}^{\mathrm{sca}}$. Unless otherwise stated, the SNR for this noise is $25$dB and $N^{\mathrm{T}} = N^{\mathrm {R}} =32$ and $f=100$ MHz in all experiments. The quality of reconstruction is measured using
\begin{equation} 
err={{{\| {{{\bar{\tau}}}}-{{{\bar{\tau }}}^{\mathrm{ref}}}\|}_{2}}}/{{{\| {{{\bar{\tau }}}^{\mathrm{ref}}}\|}_{2}}}
\end{equation}  
where ${{\bar{\tau}}}$ stores the samples of the reconstructed contrast. 
 
\subsection{Closely Spaced Point Like Targets}
\begin{figure}[t!]
\centering
\subfloat[]{\includegraphics[width=0.35\columnwidth]{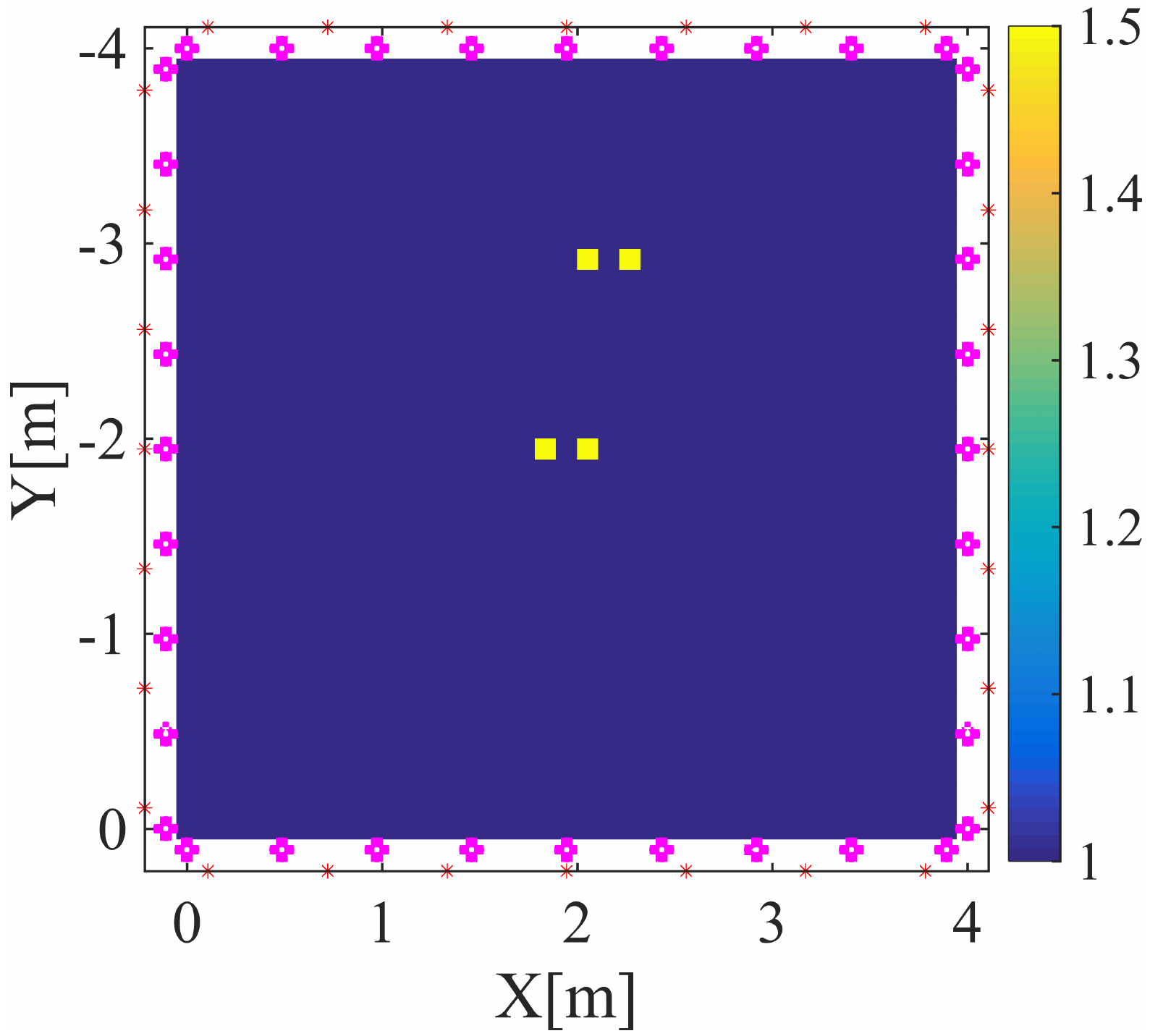}}
\subfloat[]{\includegraphics[width=0.35\columnwidth]{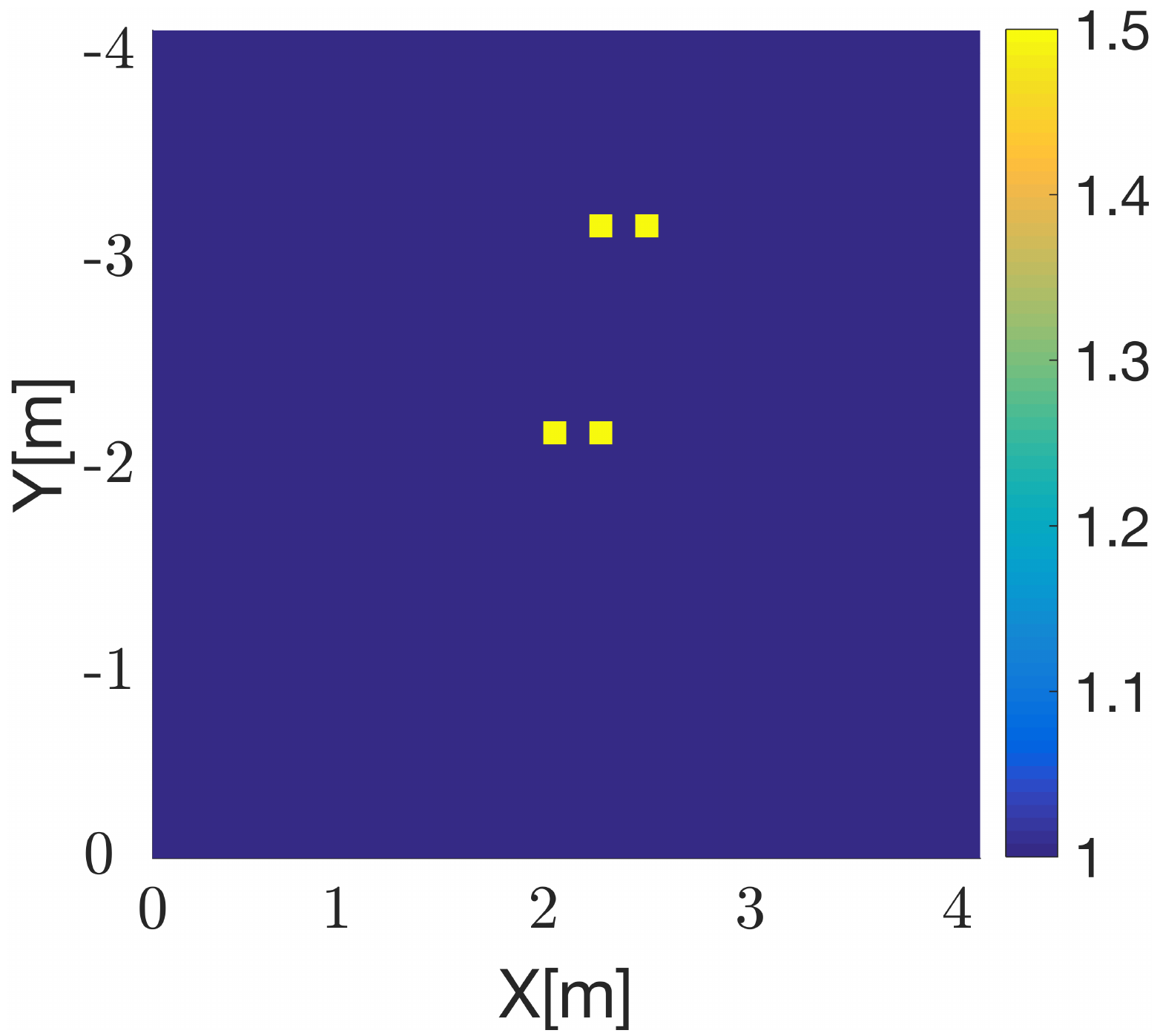}}\\
\subfloat[]{\includegraphics[width=0.6\columnwidth]{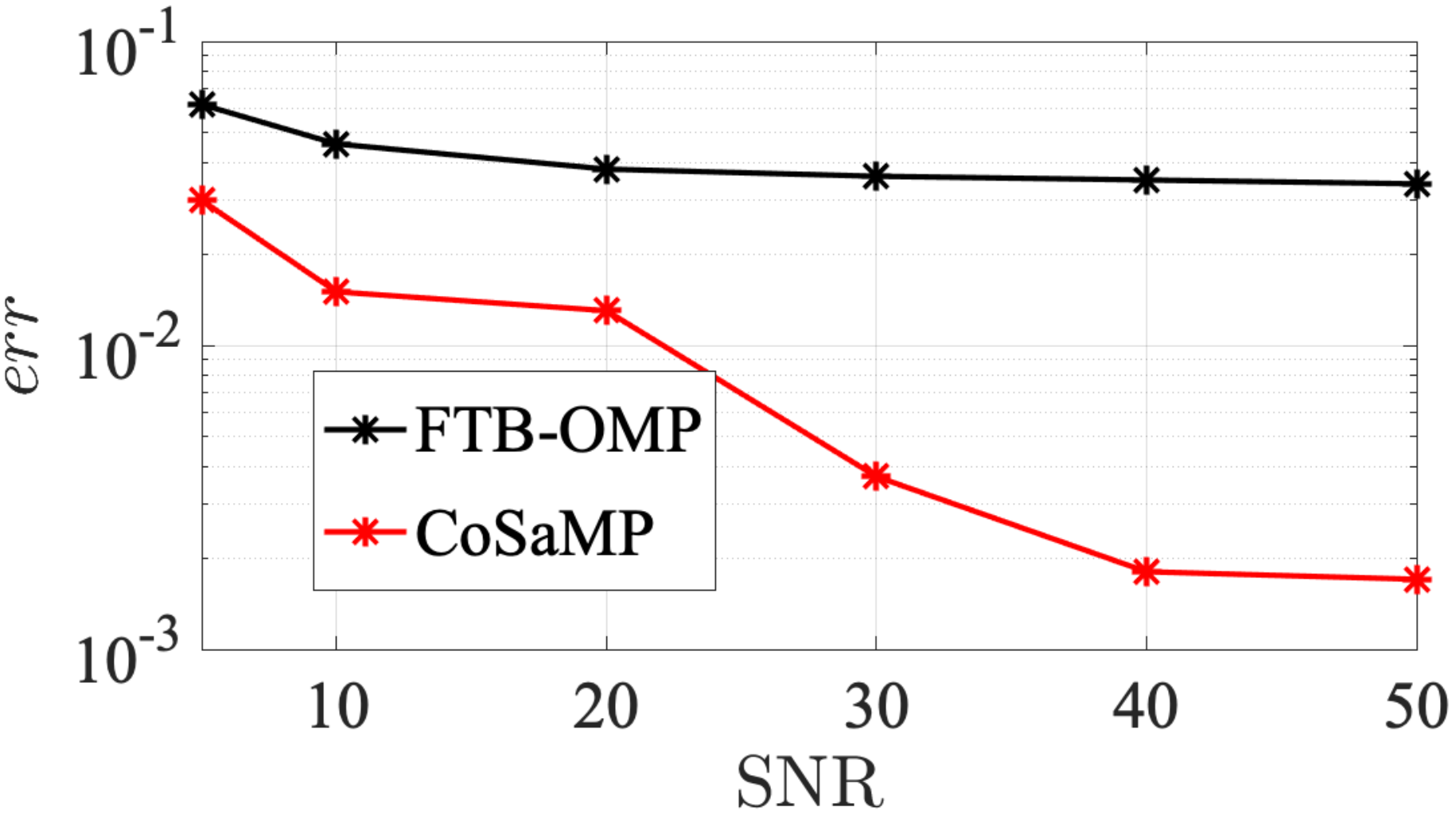}}
\caption{(a) Investigation domain ${{\bar{\tau }}^{\mathrm{ref}}}$ and transmitter-receiver configuration for two closely spaced point like targets. (b) $\bar \tau$ reconstructed using CoSaMP with $30$dB noise in $\bar{E}^{\mathrm{meas}}$. (c) $err$ in $\bar \tau$ reconstructed using CoSaMP and FTB-OMP versus six different levels of noise in $\bar{E}^{\mathrm{meas}}$.}
\label{fig3}
\end{figure}

The investigation domain ${{\bar{\tau }}^{\mathrm{ref}}}$ with $N=1369$ and $k=4$ and the transmitter-receiver configuration with $N^{\mathrm{T}}=30$, $N^{\mathrm {R}}=50$, and $f=300$ MHz are shown in Fig.~\ref{fig3}(a). Note that this example is taken from~\cite{2015ftb_omp}, where the FTB-OMP algorithm is introduced. Twelve reconstructions are carried out using CoSaMP and FTB-OMP for six different SNR values of the noise in $\bar{E}^{\mathrm{meas}}$. Fig.~\ref{fig3}(c) plots $err$ versus the SNR value. It is clear from the figure that CoSaMP is more accurate than FTB-OMP over the whole range of SNR values considered for this example. Fig.~\ref{fig3}(b) shows the reconstruction obtained using CoSaMP with $30$dB noise in $\bar{E}^{\mathrm{meas}}$. Note that, for this example, $k$ is not predicted using an ANN, instead its actual value is provided to CoSaMP. 

\subsection{Closely Spaced Cylinders}

\begin{figure}[t!]
\centering
\subfloat[]{\includegraphics[width=0.34\columnwidth]{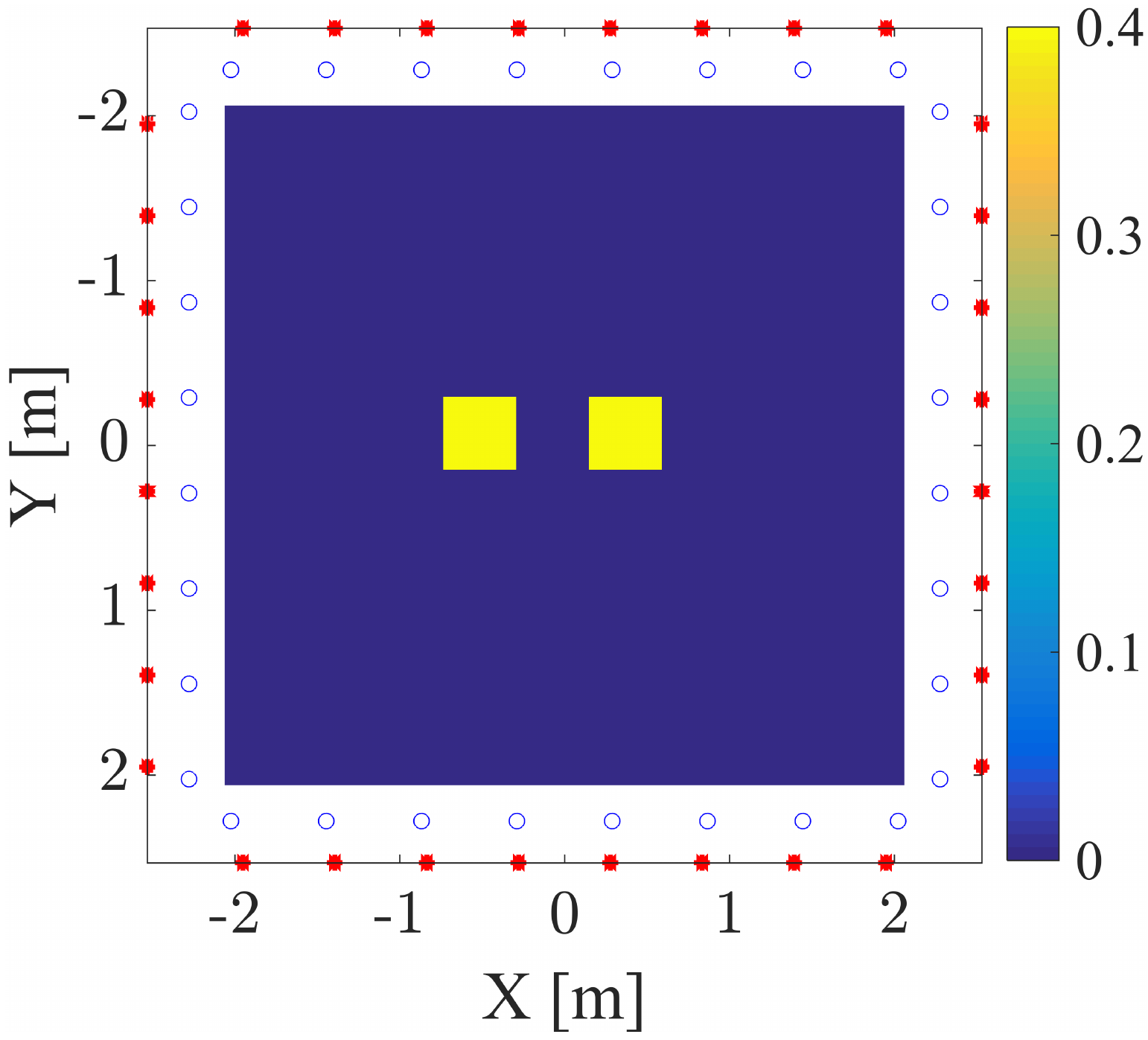}}
\subfloat[]{\includegraphics[width=0.33\columnwidth]{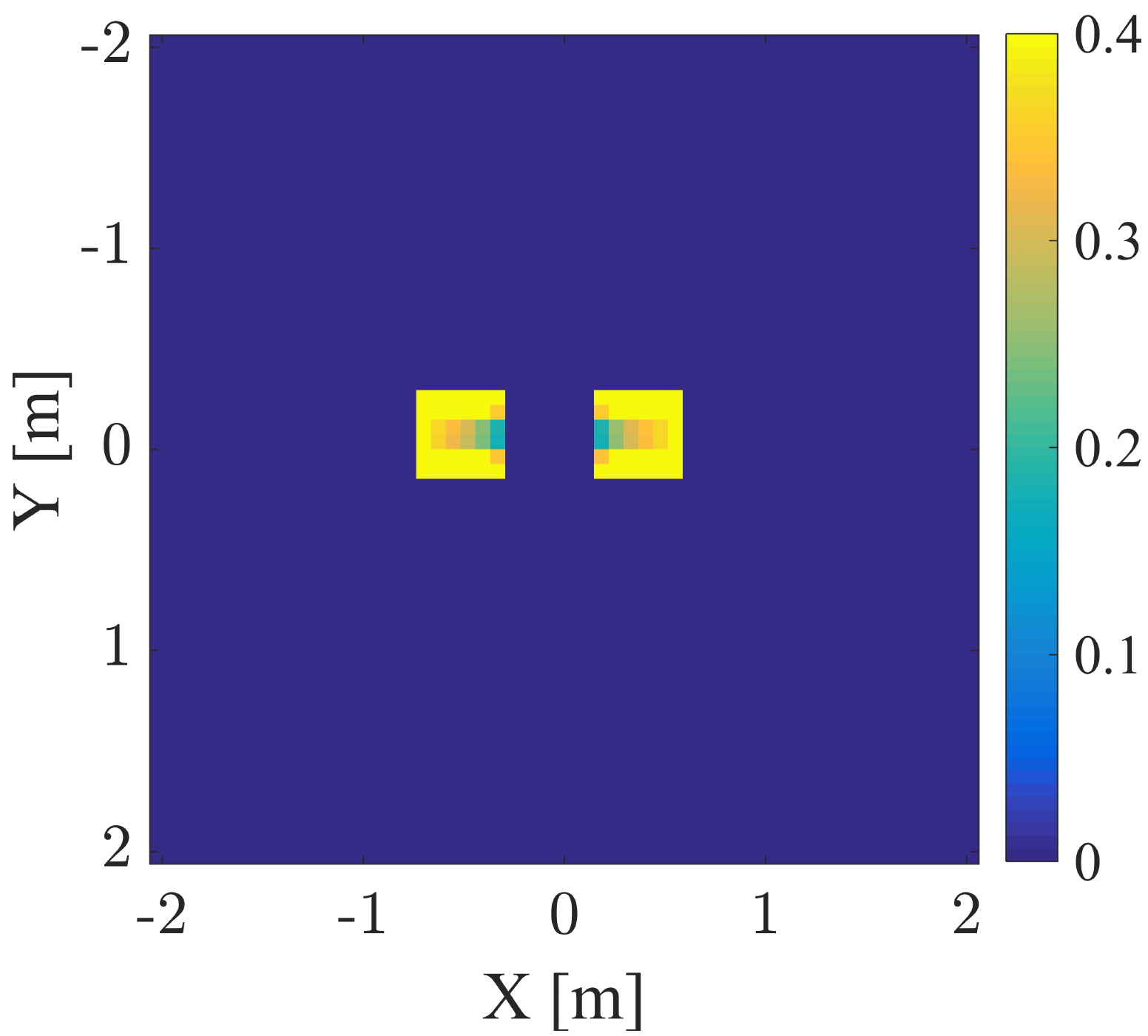}}
\subfloat[]{\includegraphics[width=0.33\columnwidth]{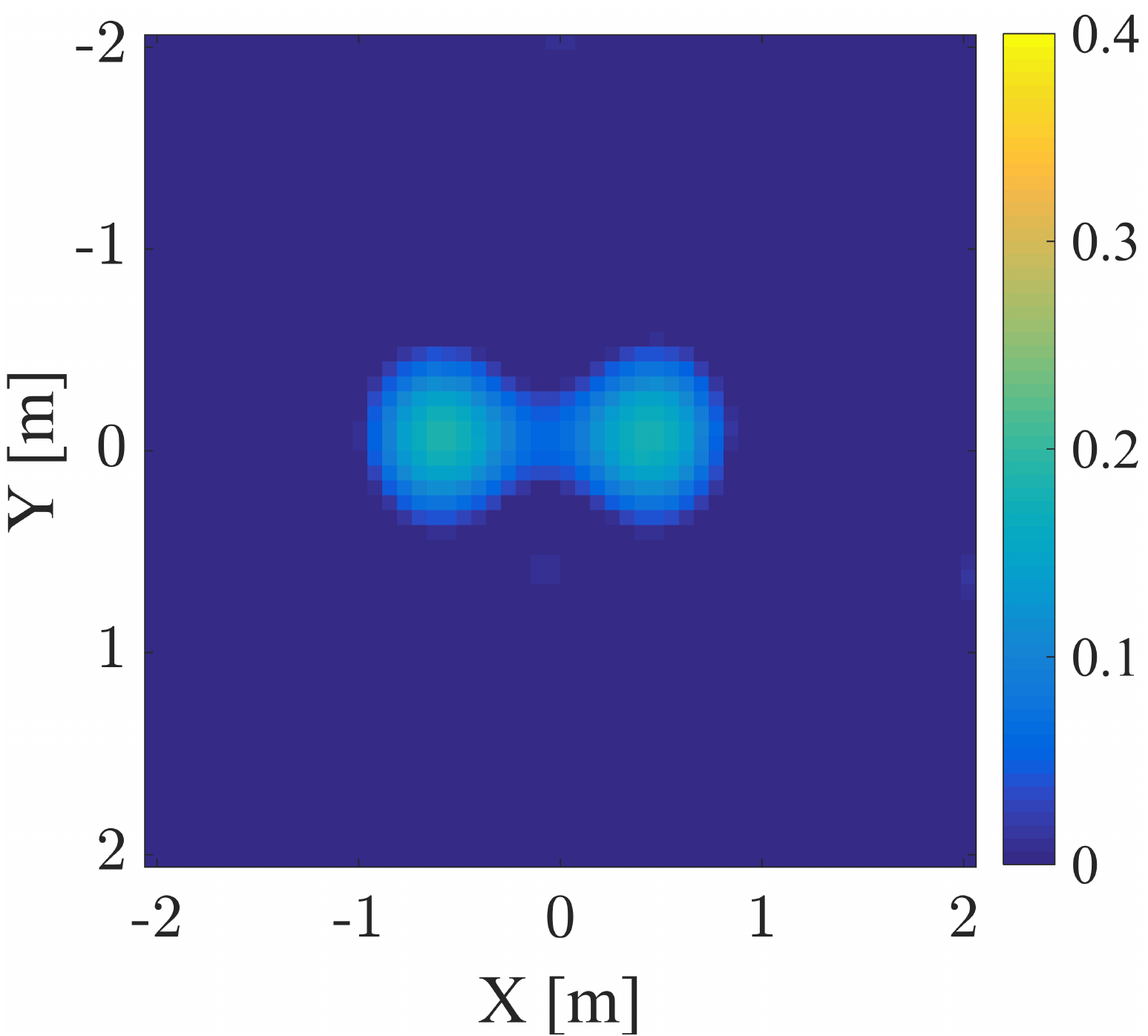}}\\
\subfloat[]{\includegraphics[width=0.5\columnwidth]{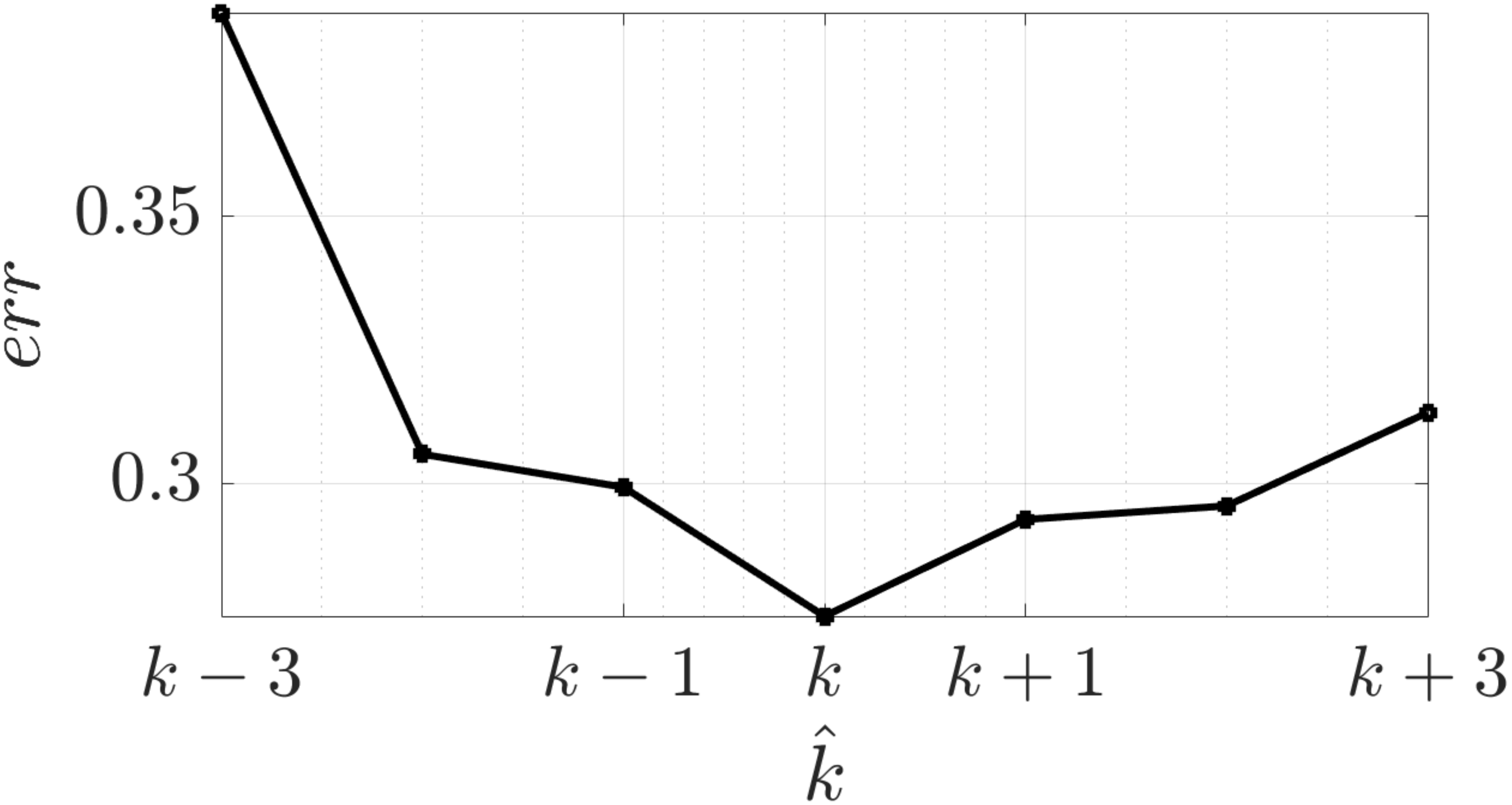}}
\subfloat[]{\includegraphics[width=0.5\columnwidth]{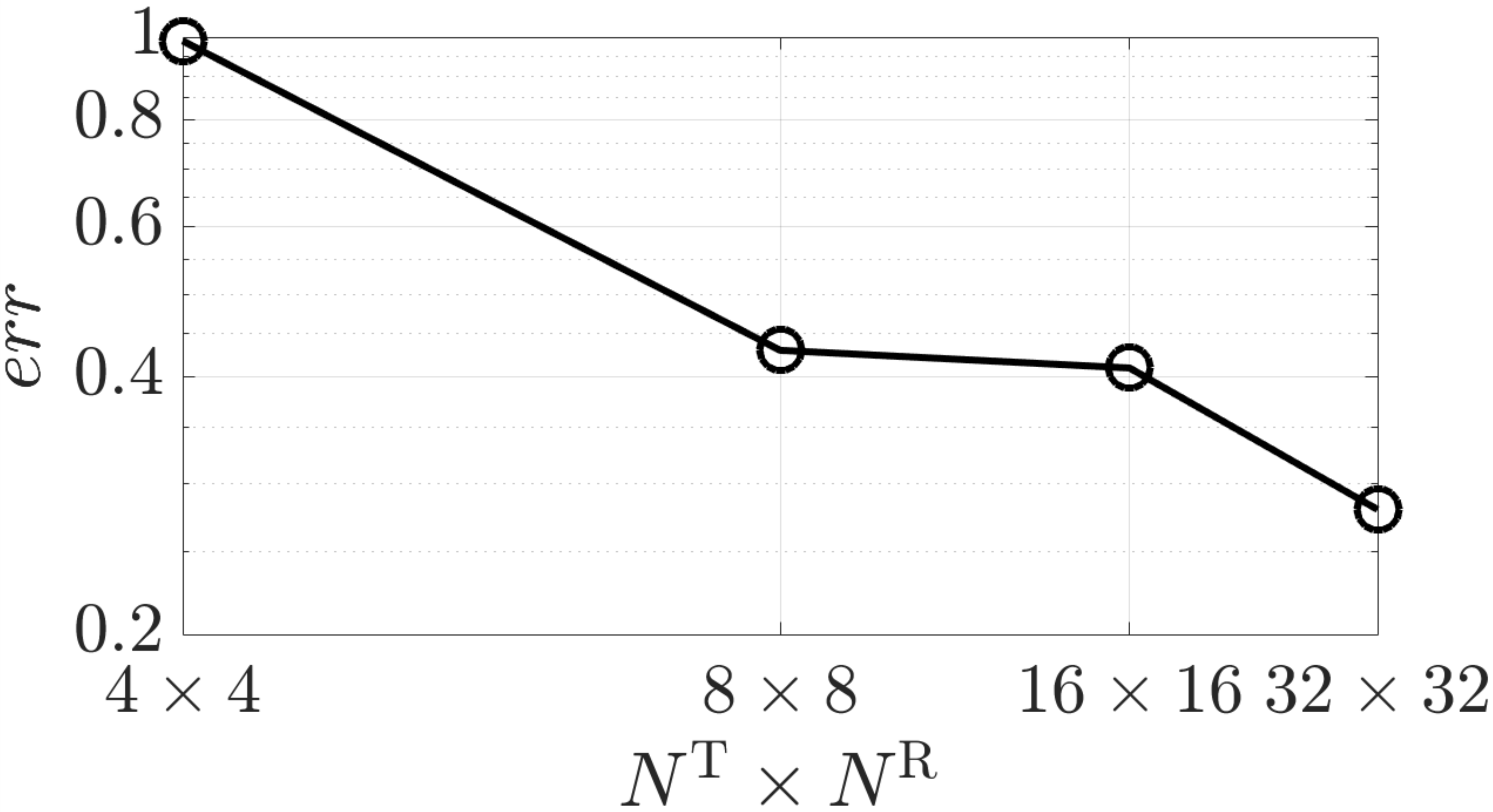}}
\caption{(a) Investigation domain ${{\bar{\tau }}^{\mathrm{ref}}}$ and transmitter-receiver configuration for two closely spaced cylinders. $\bar \tau$ reconstructed using (b) CoSaMP with $\hat k=k$ (as estimated by ANN-1) and (c) Born approximation with soft-thresholding. (d) $err$ in $\bar \tau$ reconstructed using CoSaMP versus different values of $\hat k$. (e) $err$ in $\bar \tau$ reconstructed using CoSaMP with $\hat k=k$ (as estimated by ANN-4, ANN-3, ANN-2 and ANN-1) versus $N^{\rm T} \times N^{\rm R}$.}
\label{fig4}
\end{figure}

The investigation domain ${{\bar{\tau }}^{\mathrm{ref}}}$ with $N=3136$ and $k=72$ is shown in Fig.~\ref{fig4}(a). ANN-1 estimates $k$ exactly, i.e., $\hat k=k$. Fig.~\ref{fig4}(b) and (c) show reconstructions obtained using CoSaMP and the Born approximation with soft thresholding \cite{desmal2014shrinkage}, respectively. Clearly the image produced by CoSaMP is sharper and more accurate with $err = 28\%$. It is discussed in Section II-C that the maximum value of $|\hat{k}-k|$ for all test examples is $3$ (even though for this specific example $\hat{k}=k$). Therefore, CoSaMP is executed with $\bar k \in \{k-3, k-2 \ldots k+2, k+3\}$ just to demonstrate the effect of $\bar k$ on the accuracy of reconstruction. As expected, as shown in Fig.~\ref{fig4}(d), the accuracy is highest when $\hat{k}=k$. 

Next, the effect of using smaller $N^{\rm T}$ and $N^{\rm R}$ on the accuracy of the solution is demonstrated. The reconstruction of the investigation domain described above is carried out for $\{N^{\rm T}, N^{\rm R} \} $ = $ \{16, 16\}$, $\{8,8\}$, and $ \{4, 4\}$ (in addition to $\{N^{\rm T}, N^{\rm R} \} = \{32,32 \}$ case above). The transmitters and receivers surround the investigation domain. Just like ANN-1, ANN-2, ANN-3, and ANN-4 estimate $k$ exactly for these three configurations, respectively. Fig.~\ref{fig4}(e) plots the reconstruction error versus $N^{\rm T} \times N^{\rm R}$. The figure shows that the accuracy of reconstruction degrades significantly for $N^{\rm T}=4$ and $N^{\rm R}=4$.  This demonstrates that the CoSaMP algorithm might fail to achieve a good solution when the number of measurements is too small (even though the ANN predicts the exact $k$).

\subsection{Austria}
\begin{figure}[!t]
\centering
\subfloat[]{\includegraphics[width=0.34\columnwidth]{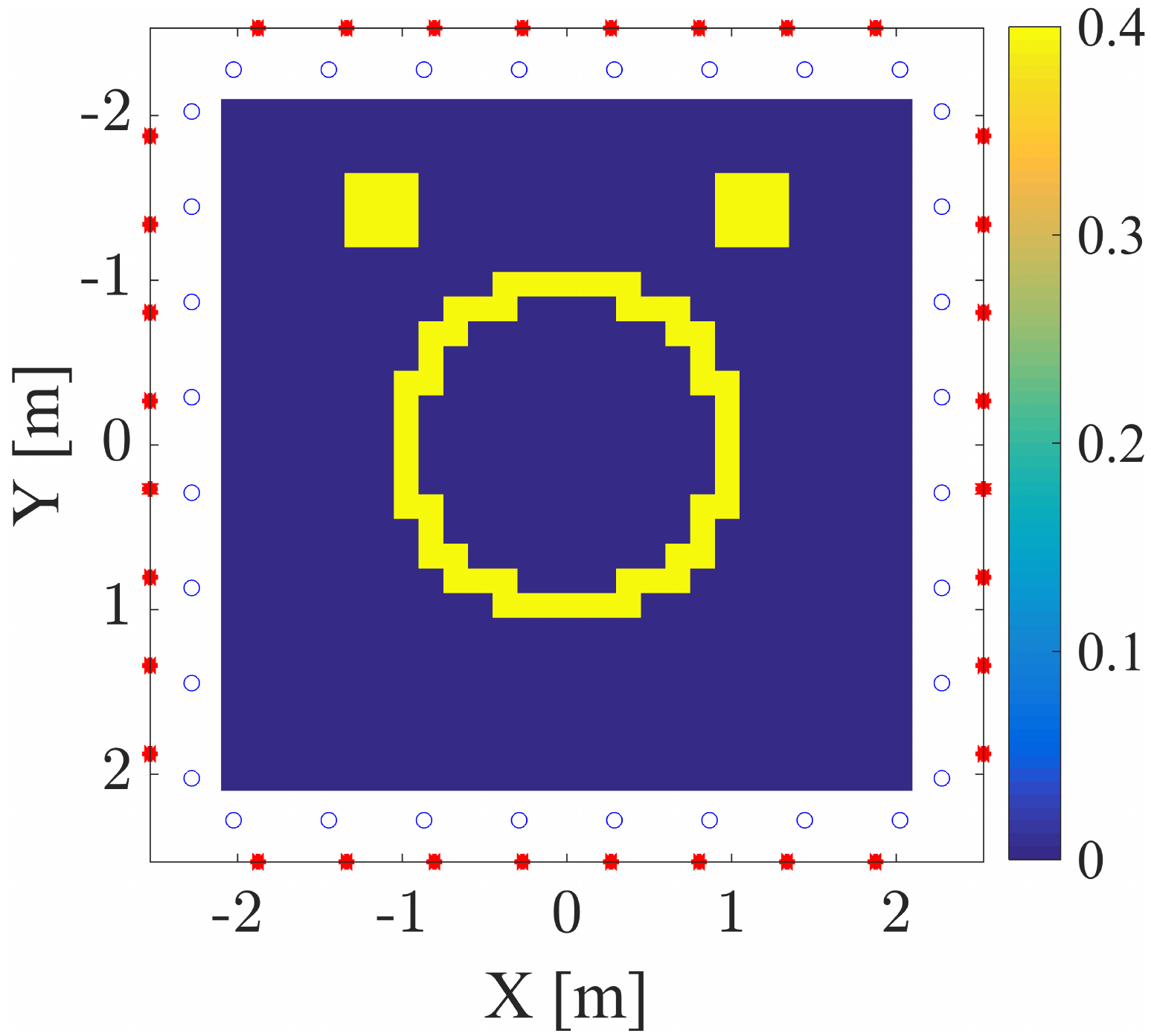}}
\subfloat[]{\includegraphics[width=0.33\columnwidth]{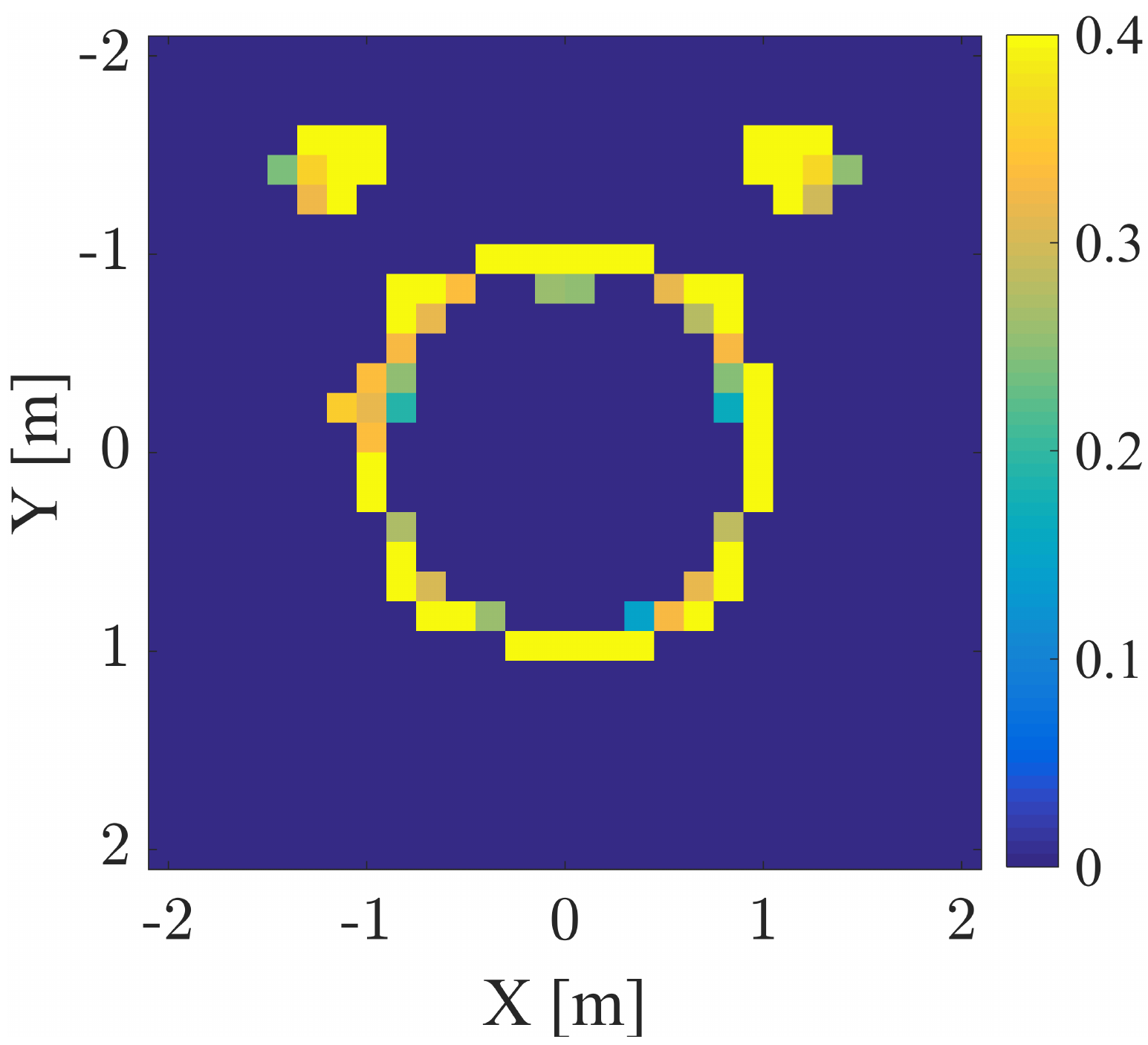}}
\subfloat[]{\includegraphics[width=0.33\columnwidth]{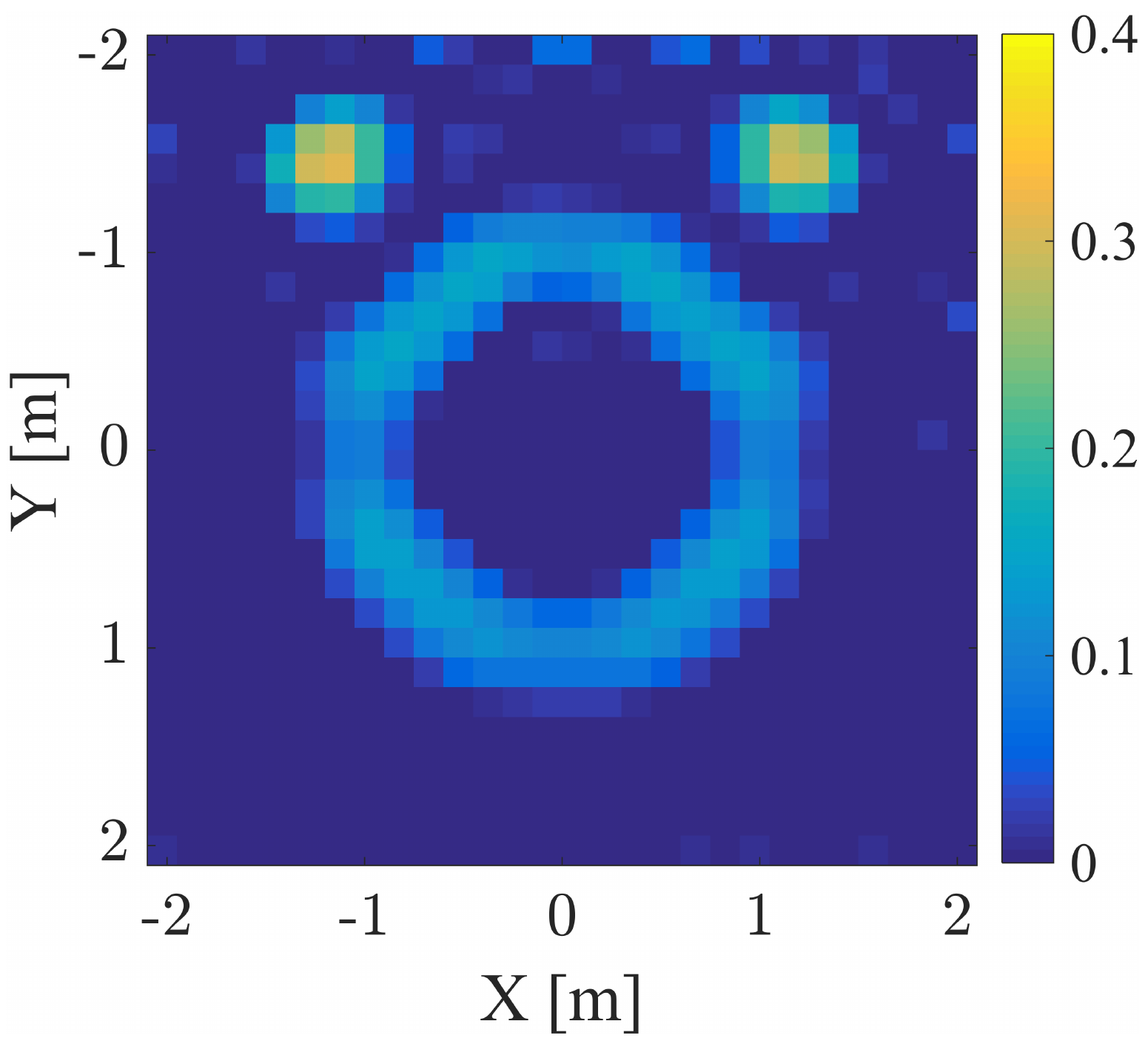}}\\
\subfloat[]{\includegraphics[width=0.5\columnwidth]{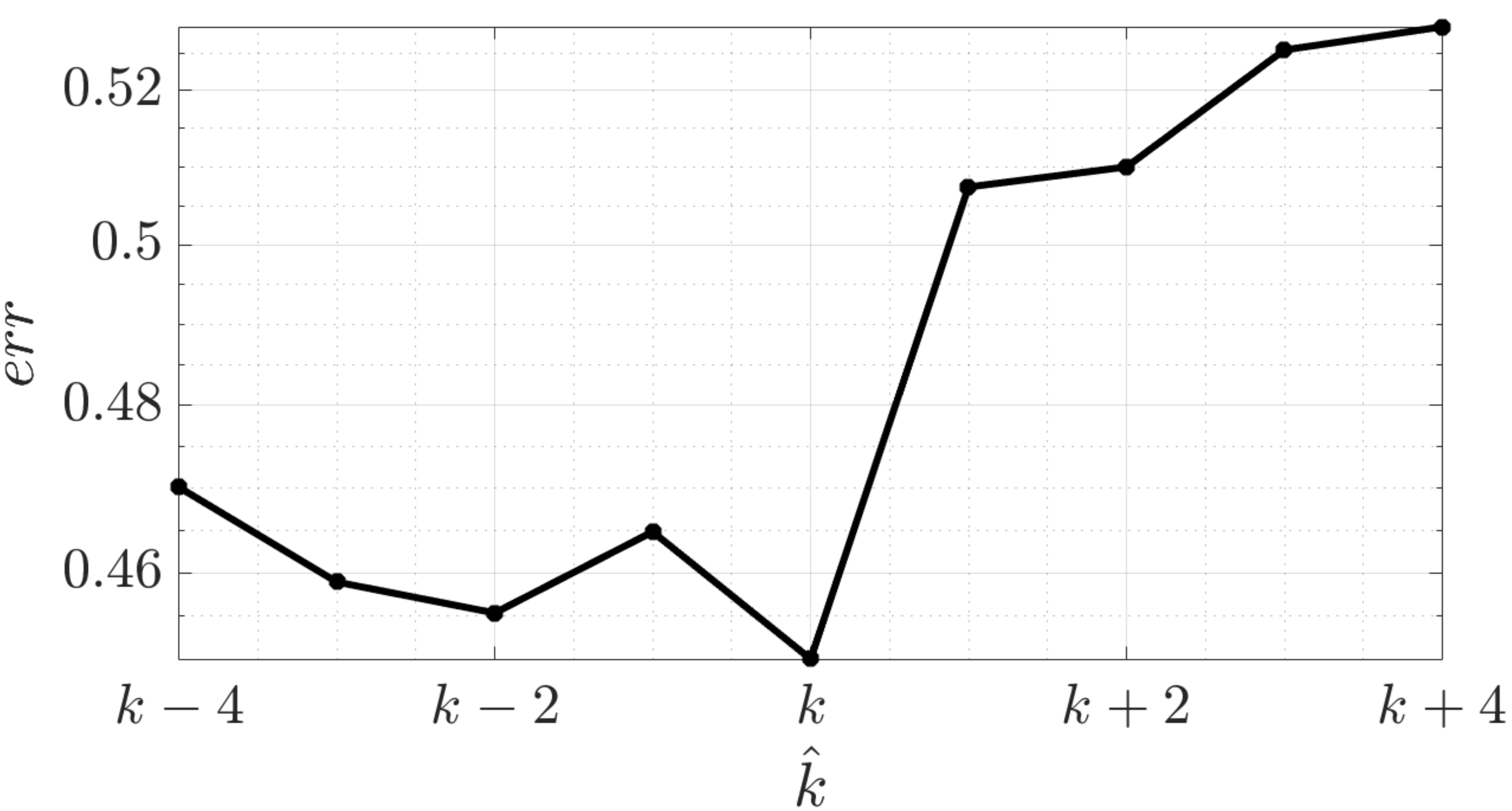}}\\
\caption{ (a) Investigation domain ${{\bar{\tau }}^{\mathrm{ref}}}$ and transmitter-receiver configuration for Austria. $\bar \tau$ reconstructed using (b)  CoSaMP with $\hat k=k$ (as estimated by ANN-1) and (c) Born approximation with soft-thresholding. (d) $err$ in $\bar \tau$ reconstructed using CoSaMP versus different values of $\hat k$.}
\label{fig5}
\end{figure}

The third example is the well-known Austria profile~\cite{van2001contrast}. The investigation domain ${{\bar{\tau }}^{\mathrm{ref}}}$ with $N=784$ and $k=66$ is shown in Fig.~\ref{fig5}(a). ANN-1 estimates $k$ exactly, i.e., $\hat k=k$. Reconstructions obtained using CoSaMP and the Born approximation with soft thresholding \cite{desmal2014shrinkage}, are shown in Fig.~\ref{fig5}(b) and (c) , respectively. CoSaMP is more accurate with $err = 43\%$. Similar to the previous example, just to demonstrate the effect of $\hat k$ on the accuracy of reconstruction, CoSaMP is executed with $\bar k \in \{k-4, k-3 \ldots k+3, k+4\}$ (even though the ANN predicts $\hat{k}=k$) and Fig.~\ref{fig5}(d)  shows that reconstruction accuracy degrades as $|\hat{k}-k|$ increases.

\subsection{L-Shaped Object}

\begin{figure}[t!]
\centering
\subfloat[]{\includegraphics[width=0.35\columnwidth]{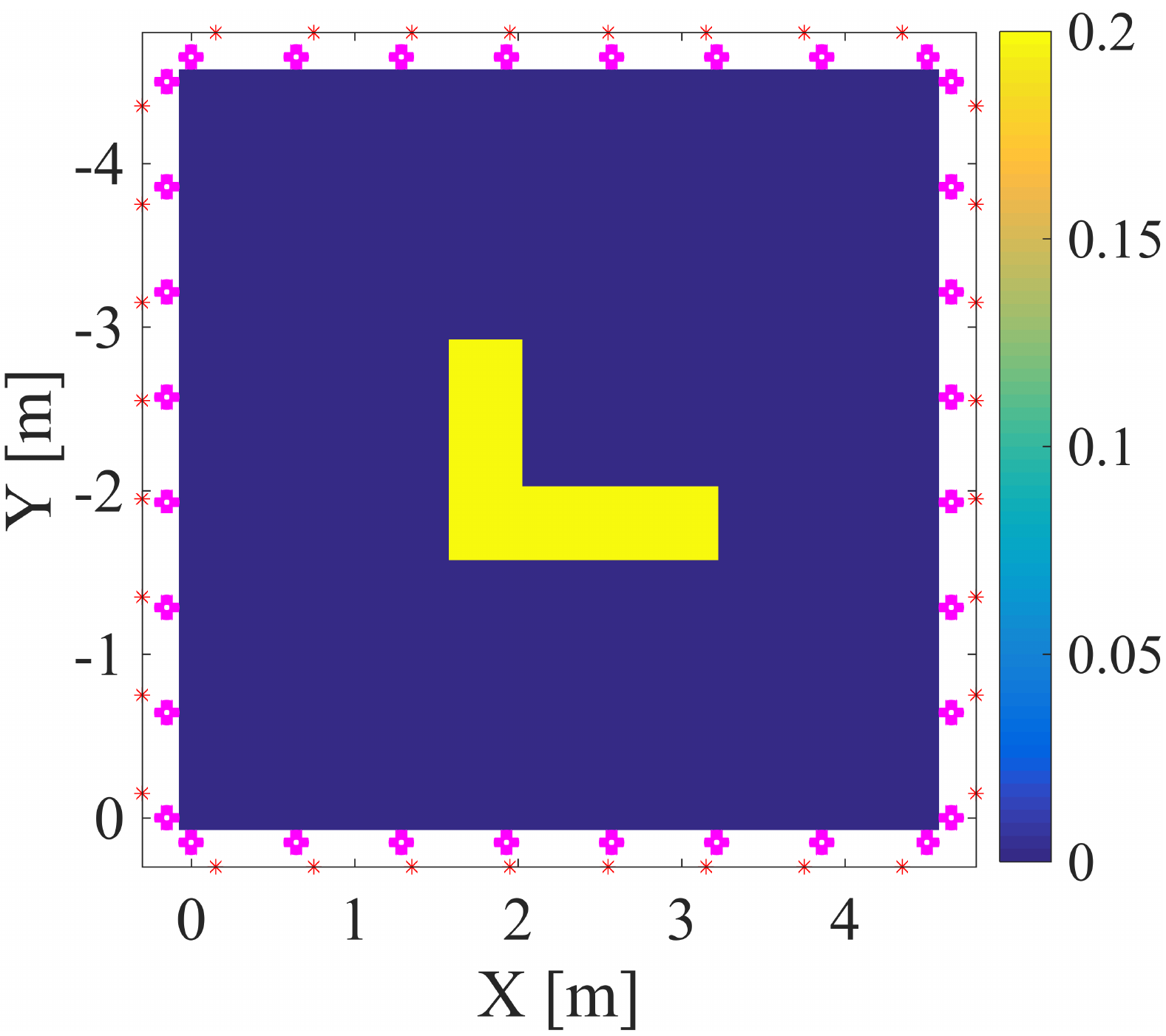}}
\subfloat[]{\includegraphics[width=0.35\columnwidth]{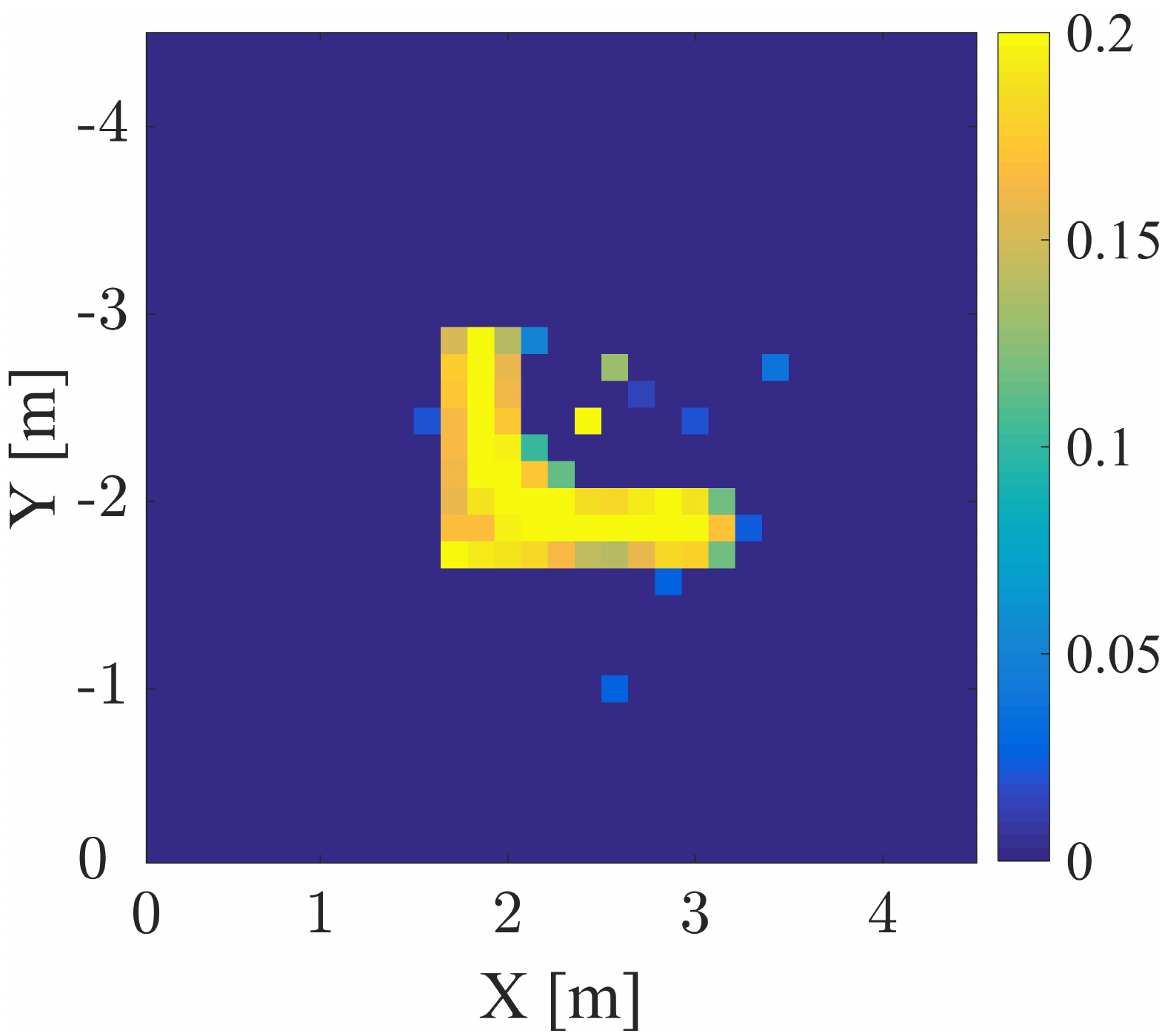}}
\caption{(a) Investigation domain ${{\bar{\tau }}^{\mathrm{ref}}}$ and transmitter-receiver configuration for L-shaped object. (b) $\bar \tau$ reconstructed using CoSaMP.}
\label{fig6}
\end{figure}

The last example is an L-shaped object. The investigation domain ${{\bar{\tau }}^{\mathrm{ref}}}$ with $N=961$ and $k=65$ is shown in Fig.~\ref{fig6}(a). CoSaMP reconstructs this object with $err=31.6\%$ as shown in Fig.~\ref{fig6}(b).

\section{Conclusion}

A greedy algorithm is used together with a simple ANN for efficient and accurate EM imaging of 2D sparse investigation domains. To enable the application of CoSaMP to solving the EM inverse scattering problem (i) a simple ANN is used to predict the number of non-zero contrast samples in the investigation domain and (ii) a constant is added to the diagonal entries of the scattering matrix (resulting in a Tikhonov-type regularization), which ensures that the RIP condition is satisfied. The resulting EM inversion scheme is computationally efficient since it calls for solution of a smaller least squares problem on a reduced set determined by the number of non-zero contrast samples. 
Numerical results, which demonstrate that the proposed scheme produces more accurate and sharper images than Born approximation with soft thresholding, are provided.

\section*{Acknowledgment}
The authors would like to thank the King Abdullah University of Science and Technology (KAUST) Supercomputing Laboratory (KSL) for providing the required computational resources.


\end{document}